\documentclass[twocolumn]{aastex6}

\bibpunct{(}{)}{;}{a}{}{,}
\begin{document}

\title{Recovering the properties of high redshift galaxies with different \emph{JWST} broad-band filters}
\author{L. Bisigello\altaffilmark{1,2}, K. I. Caputi\altaffilmark{1}, L. Colina\altaffilmark{3}, O. Le F\`evre\altaffilmark{4}, H. U. N\o rgaard-Nielsen\altaffilmark{5}, P. G. P\'erez-Gonz\'alez\altaffilmark{6},  P. van der Werf\altaffilmark{7}, O. Ilbert\altaffilmark{4}, N. Grogin\altaffilmark{8}, A. Koekemoer\altaffilmark{8}
}

\altaffiltext{1}{Kapteyn Astronomical Institute, University of Groningen, P.O. Box 800, 9700 AV, Groningen, The Netherlands.}
\altaffiltext{2}{SRON Netherlands Institute for Space Research, 9747 AD, Groningen, The Netherlands. 
\\ Email:  bisigello@astro.rug.nl}
\altaffiltext{3}{Centro de Astrobiolog\'{\i}a, Departamento de Astrof\'{\i}sica, CSIC-INTA, Cra. de Ajalvir km.4, 28850 - Torrej\'on de Ardoz, Madrid, Spain}
\altaffiltext{4}{Aix Marseille Universit\'e, CNRS, LAM (Laboratoire d'Astrophysique de Marseille), UMR 7326, 13388, Marseille, France}
\altaffiltext{5}{National Space Institute (DTU Space), Technical University of Denmark, Elektrovej, DK-2800 Kgs. Lyngby, Denmark}
\altaffiltext{6}{Departamento de Astrof\'{\i}sica, Facultad de CC. F\'{\i}sicas, Universidad Complutense de Madrid, E-28040 Madrid, Spain; Associate Astronomer at Steward Observatory, The University of Arizona.}
\altaffiltext{7}{Sterrewacht Leiden, Leiden University, PO Box 9513, 2300 RA, Leiden, The Netherlands}
\altaffiltext{8}{Space Telescope Science Institute, 3700 San Martin Drive, Baltimore, MD 21218, USA}

\shorttitle{Bisigello et al.: High-z galaxy properties with {\em JWST} filters}

\shortauthors{Bisigello et al.}

\begin{abstract}
Imaging with the \emph{James Webb Space Telescope (JWST)} will allow for observing the bulk of distant galaxies at the epoch of reionisation. The recovery of their properties, such as age, color excess E(B-V), specific star formation rate (sSFR) and stellar mass, will mostly rely on spectral energy distribution fitting, based on the data provided by \emph{JWST}'s two imager cameras, namely the Near Infrared Camera (NIRCam) and the Mid Infrared Imager (MIRI). In this work we analyze the effect of choosing different combinations of NIRCam and MIRI broad-band filters, from 0.6 $\mu$m to 7.7 $\mu$m, on the recovery of these galaxy properties. We performed our tests on a sample of 1542 simulated galaxies, with known input properties, at $z=7-10$. We found that, with only 8 NIRCam broad-bands, we can recover the galaxy age within 0.1 Gyr and the color excess within 0.06 mag for 70$\%$ of the galaxies. Besides, the stellar masses
and sSFR are recovered within 0.2 and 0.3 dex, respectively, at z=7-9.  Instead, at z=10, no NIRCam band traces purely the $\lambda> 4000\rm\AA$ regime and the percentage of outliers in stellar mass (sSFR) increases by $>20\%$ ($>90\%$), in comparison to z$=$9. The MIRI F560W and F770W bands are crucial to improve the stellar mass and the sSFR estimation at z$=$10. When nebular emission lines are present, deriving correct galaxy properties is challenging, at any redshift and with any band combination. In particular, the stellar mass is systematically overestimated in up to 0.3 dex on average with NIRCam data alone and including MIRI observations improves only marginally the estimation. 
\end{abstract}


\section{Introduction} \label{sec:intro}

Fitting the spectral energy distribution (SED) of galaxies from broad band photometry allows for deriving a variety of galaxy properties, such as photometric redshifts, stellar masses, color excess, ages and metallicities, with different degrees of precision, depending on the available data and used models. Particularly, SED fitting is a powerful tool to study the properties of galaxies up to high redshifts, due to the fainter fluxes usually observable with photometry respect to spectroscopy. \par
The stellar mass is one of the most important quantities to derive for a galaxy, because it correlates with a large number of global properties, such as star formation rate \citep[SFR; e.g.][]{Brinchmann2004,Noeske2007,Rodighiero2011} and metallicity \citep[e.g. ][]{Tremonti2004,Erb2006,Maiolino2008}, and it is central in galaxy evolution. Fortunately, once the redshift is well determined, stellar masses are one of the most robust parameters estimated from the SED fitting  \citep[e.g.][]{Caputi2015}, while others parameters, such as age and color excess, are generally more difficult to estimate, given the degeneracy between them. However, the inclusion of nebular emission can have a large effect on the stellar masses \citep[e.g.][]{Stark2013,Santini2015}. At low and intermediate redshifts, the emission line equivalent widths are relatively low for most galaxies and, therefore, nebular emission can be safely ignored in stellar mass calculations. However, at higher redshifts, young galaxies are more common and, therefore, the effect of nebular emission is expected to be more important. \par
In addition, stellar masses are derived multiplying the galaxy luminosity with a mass-to-light ratio, thus, the mass estimation depends also on the wavelengths of the available observations. Observations at rest-frame $\lambda>4000\rm\AA$, and ideally at $\lambda>1 \mu m$, are necessary to derive stellar masses, because the mass-to-light ratio at these wavelengths are relatively insensitive to age and color excess. Therefore, in order to derive accurate stellar masses in a wide range of redshifts, it is necessary to observe in the near-IR at low and intermediate redshifts, and mid-IR at high redshifts.\par
 The \emph{James Webb Space Telescope} (\emph{JWST}\footnote{http://www.jwst.nasa.gov},\citealt{Gardner2009}) is a premier infrared space observatory for the next decade with a 6.5-meter primary mirror. It has four scientific instruments on board with imaging, spectroscopic and coronographic modes, covering a wide range of wavelengths from the visible to the mid-IR (0.6-28 $\mu$m) with superb sensitivity and resolution. In particular, the two imaging cameras, namely the Near Infrared Camera \citep[NIRCam;][]{Rieke2005} and the Mid Infrared Instrument \citep[MIRI;][]{Rieke2015,Wright2015}, will be used to carry out deep blank-field imaging surveys to study how galaxies assemble and evolve since early cosmic times. For these surveys, SED-fitting analysis will be performed to derive galaxy properties, such as galaxy stellar mass, ages and color excesss, therefore, it is important to understand the effect of different filter combination choices on the ability to estimate these quantities.\par
NIRCam\footnote{http://www.stsci.edu/jwst/instruments/nircam} is equipped with 8 broad-band filters (i.e. F070W, F090W, F115W, F150W, F200W, F277W, F356W and F444W) together with a number of medium and narrow-band filters, covering the range from 0.6 to 5 $\mu$m. The MIRI imager\footnote{http://www.stsci.edu/jwst/instruments/miri} is complementary to NIRCam in wavelengths, covering between 5 and 28 $\mu$m with nine broad-band filters, to observe both the redshifted stellar light at high-z and the hot dust radiation at low-z. Only the two shortest wavelengths MIRI filters are considered in this work (i.e. F560W and F770W), because they are the most sensitive ones and, therefore, the ones that will be mostly used for high-z galaxy studies. \par
Because of the complementary wavelength range covered by NIRCam and MIRI, they should ideally be both included in deep galaxy surveys, particularly for z$\geqslant$7 where NIRCam observes the rest-frame UV/optical of the SED while MIRI covers the rest-frame optical/near-IR. However, because of different detector technology used at near- and mid-IR wavelengths, MIRI is less sensitive that NIRCam, making observations with this instrument more time-expensive. In this work, we address this issue, analyzing the galaxy property estimation with different NIRCam and MIRI filter combinations and for different spectral templates. We treated all situations on an equal basis without judging how common different spectral types are expected to be, in order to leave the reader decide the optimal approach for each science case. \par
In this work we aimed to evaluate the effect of different \emph{JWST} broad-band filter combinations on deriving secure galaxy properties, i.e. stellar masses, color excess and age, for high-z galaxies.  We applied our test on a sample of simulated galaxies at $z=7,8,9$ and 10 derived from \cite{Bisigello2016} (hereafter B16), where the photometric redshift recovery is analyzed in detail for the same combinations of \emph{JWST} broad-band filters. \par
 The structure of the paper is the following. In section \ref{sec:sample} we describe the analyzed sample, the obtention of the photometry in the pertinent NIRCam and MIRI bands and the galaxy properties derivation. We present our results at different redshifts in section \ref{sec:results}: in particular, we analyze stellar masses, age, color excess and specific star formation rate (sSFR). In section \ref{sec:conclusions} we summarize our main findings and conclusions. Throughout this paper, we adopt a cosmology with $H_0=$70~$\rm km \, s^{-1} \, Mpc^{-1}$, $\Omega_M=0.27$, $\Omega_\Lambda=0.73$. All magnitudes refer to the AB system \citep{Oke1983} and we considered a \cite{Chabrier2003} initial mass function (IMF). \par

\section{Sample selection and test methodology}\label{sec:sample}

\subsection{Sample construction}
\begin{figure*}[ht!]
\center{
\includegraphics[width=0.9\linewidth, keepaspectratio]{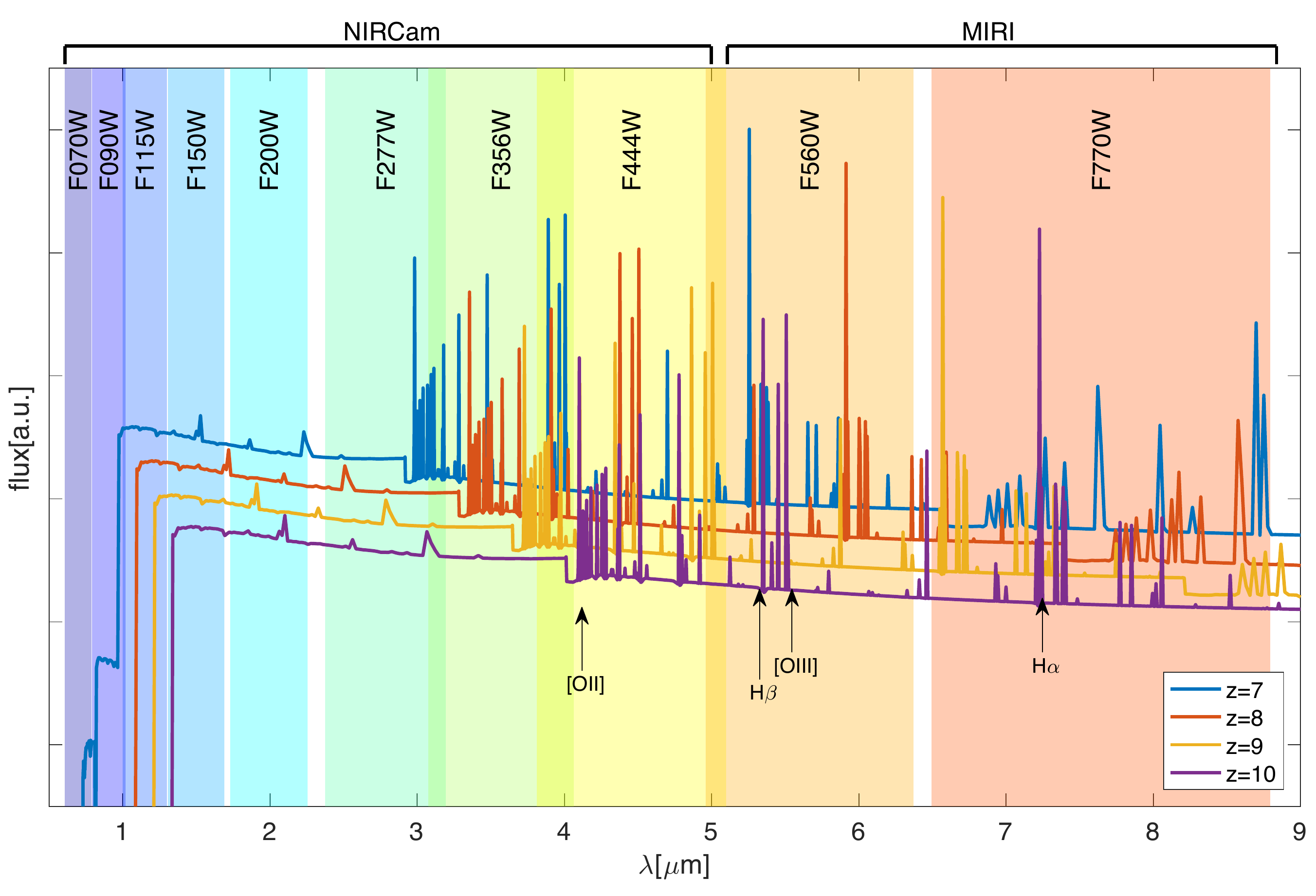}}
\caption{Four examples of SED templates from the population synthesis code \textit{Yggdrasil} with f$_{cov}=$1. The colored areas specify the wavelength range of the 8 NIRCam broad-band filters and MIRI F560W and F770W. This Figure illustrate the amount of emission lines incorporate in these SED templates and in which \textit{JWST} broad-band filters they are observed at each analysed redshift.  \label{fig:Ygg_bands}}
\end{figure*}

Our analyses is based on a sample of 1542 simulated galaxies at $z=7-10$, presented in B16.
The sample is based on simulated galaxies derived using templates from \citet[][BC03 hereafter]{B&C2003} with the manual addition of the main emission lines and templates obtained with the population synthesis code \textit{Yggdrasil} \citep{Zackrisson2011}.
The main difference between the two types of templates is the incorporation of emission lines. The BC03 models do not include emission lines, so, when galaxies are young and star-forming, we added manually the main emission lines (H$\alpha$, H$\beta$, [OIII] at 5007$\rm\AA$ and [OII] at 3727$\rm\AA$), whose rest-frame equivalent widths have been assumed to change with redshift and metallicity (see B16 for more details). In the \textit{Yggdrasil} templates, nebular emission lines are automatically incorporated together with the nebular continuum emission, when galaxies are young and the gas covering factor f$_{cov}>$0. For this second family of templates, rest-frame equivalent widths change with metallicity rather than redshift. In Fig. \ref{fig:Ygg_bands}, there is an example of an \text{Yggdrasil} SED template with f$_{cov}=1$ at the four considered redshifts (7, 8, 9 and10), to show the amount of emission lines that contaminate the \textit{JWST} broad-band filters analysed on this work at each redshift.\par
A second difference between the two types of template is the star formation history. In particular, the BC03 models have declining star formation histories, while \textit{Yggdrasil} templates have step function star formation histories. It is necessary to remember this difference when analyzing the sSFR estimation for the two types of templates, as it is discussed in more details in the section \ref{sec:sSFR}. An additional BC03 SED model with increasing star formation history is analyzed in Appendix \ref{sec:AppA}.\par
For both templates we considered a large range of parameters consistent with $z=7-10$ (Table \ref{tab:param}). We did not include galaxies with Population III stars among \textit{Yggdrasil} templates, as in B16, because the stellar mass is not defined for them. All templates are normalized at 29 AB mag at 1.5 $\mu$m, corresponding to the pivot wavelength of the NIRCam F150W broad-band filter.  \par
\begin{figure}[ht!]
\center{
\includegraphics[width=1\linewidth, keepaspectratio]{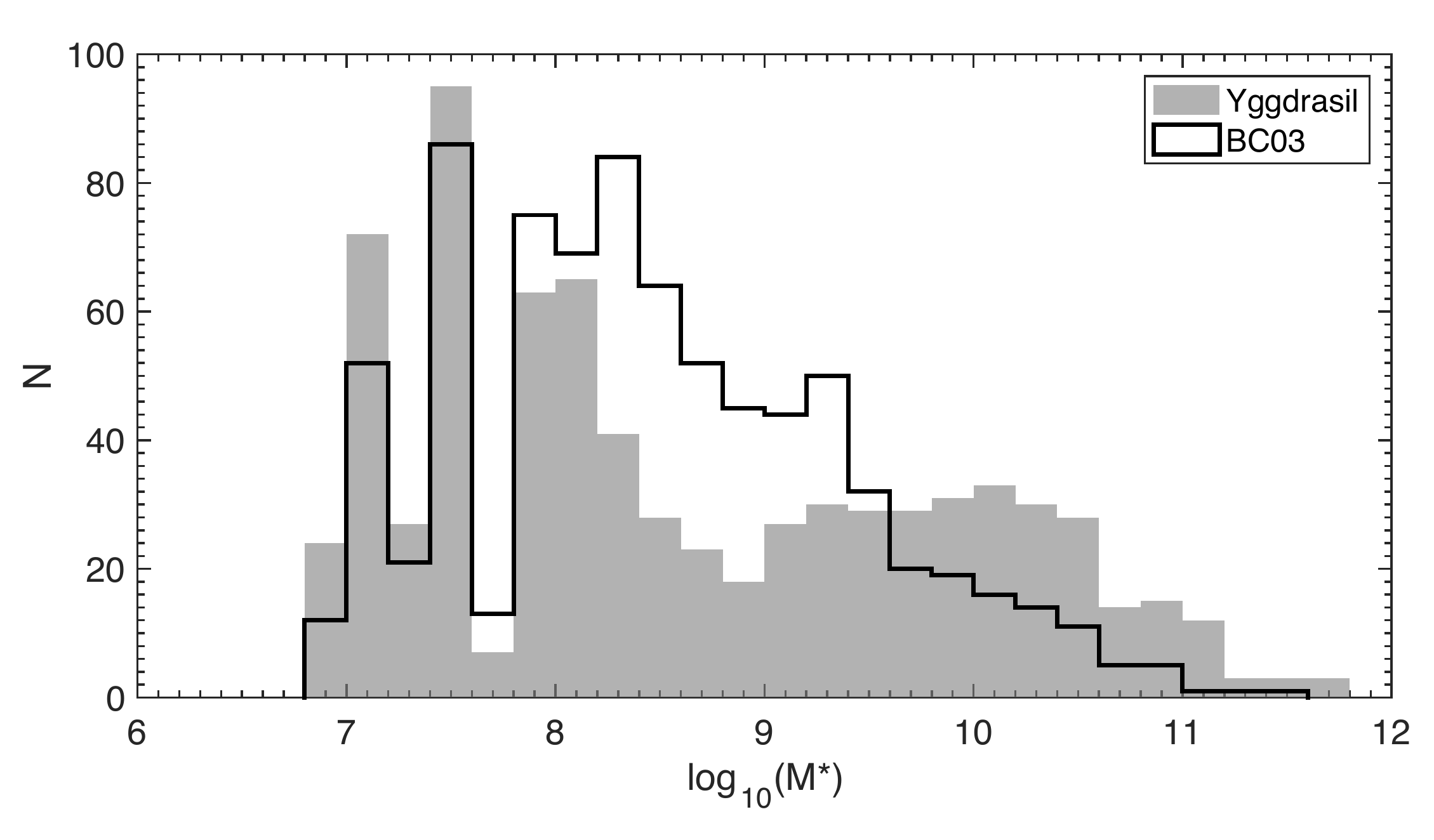}}
\caption{Input stellar masses of galaxies simulated with BC03 templates (\textit{black line}) and \emph{Yggdrasil} models (\textit{grey area}). All these stellar masses correspond to SED models normalized at 29 AB magnitude at observed 1.5 $\mu$m. We considered a \cite{Chabrier2003} IMF. \label{fig:mass3}}
\end{figure}

The full sample consists of 1542 galaxies: there are 417 SED models at each redshift 7 and 8 (216 BC03 templates and 201 \textit{Yggdrasil} templates) and 354 SED models at each redshift 9 and 10 (180 BC03 templates and 174 \textit{Yggdrasil} templates). The distribution of the stellar mass for all templates in this sample (corresponding to the adopted normalization) is shown in Fig. \ref{fig:mass3}. Differences in the two mass distributions are due to the different star formation histories used for the two types of templates and the presence of young galaxies without emission lines ($f_{cov}=0$) among the \emph{Yggdrasil} templates that are not present among BC03 templates. All results are valid also for lower masses, as long as the SED shapes and S/N values are the same. Note that our sample is not intended to emulate the distribution of galaxy SED in a real galaxy population, but rather span the different types of SED types of the galaxies that will be observed in high-z \emph{JWST} NIRCam and MIRI blank field, on an equal basis. \par

\begin{deluxetable*}{ccc}
\tablecaption{Parameter values used to create BC03 and \textit{Yggdrasil} SED models of the simulated galaxies at z$=7-10$ that are analyzed in this work. \label{tab:param}}
\tablecolumns{3}
\tablewidth{0pt}
\tablehead{
\colhead{Parameter} &
\colhead{Values (BC03)} &
\colhead{Values (Ygg)}
}
\startdata
metallicity & Z$_{\odot}$, 0.4Z$_{\odot}$, 0.2Z$_{\odot}$,0.02Z$_{\odot}\tablenotemark{a}$ & Z$_{\odot}$,0.4Z$_{\odot}$,0.2Z$_{\odot}$,0.02Z$_{\odot}$\tablenotemark{a}\\
SFH type & declining & step function \\
SFH [Gyr]& 0.01,0.1,1,10 & 0.01,0.03,0.1\\
$f_{\rm cov}$ & -- & 0\tablenotemark{b},1\\
E(B-V)\tablenotemark{c} & 0,0.1,0.25  & 0,0.1,0.25 \\
age [Gyr] & 0.01,0.05,0.2,0.4,0.6\tablenotemark{d} & 0.01,0.05,0.2,0.4,0.6\tablenotemark{d}\\
$z$ & 7,8,9,10 & 7,8,9,10\\
\enddata
\tablenotetext{a}{for this metallicity we considered only ages $t<$0.2~Gyr.}
\tablenotetext{b}{templates of old galaxies with no star formation ongoing do not change with the covering factor, so, for these galaxies, we considered only f$_{cov}=$1.}
\tablenotetext{c}{following Calzetti et al. reddening law \citep{Calzetti2000}}
\tablenotetext{d}{we considered this age only up to redshift $z=8$.}
\end{deluxetable*}

\subsection{Interpolation/extrapolation of JWST photometry}
All galaxies in the sample have mock observations for the 8 NIRCam broad bands, MIRI F560W and F770W broad bands from B16, derived by convolving each SED template with the NIRCam and MIRI filter transmission curves \citep{Meyer2004,Bouchet2015,Glasse2015}. We normalized each template at 29 AB mag at 1.5 $\mu$m, corresponding to the NIRCam F150W filter pivot wavelength, and we scaled the fiducial stellar mass of each template accordingly to this normalization. One of the major cause of errors in the stellar mass derivation is a wrong redshift estimation. In our analysis, we wanted to study independently each parameter, therefore we associated a signal-to-noise value of 10 with the flux at the F150W band, which corresponds to $\sim$8-9 hours per pointing of exposure time with low background level\footnote{derived using the JWST time exposure calculator https://jwst.etc.stsci.edu}\label{ETC}. Considering these high signal-to-noise values, the photometric redshift is well determined for $>99\%$ of the galaxies in our sample (see B16). We scaled the photometry of the other NIRCam bands considering the same integration time, but different sensitivities, according to each band. \textbf{For the MIRI bands, all fluxes correspond to the SED templates normalised to 29 mag at $1.5~ \rm \mu$m. The S/N values on the MIRI bands have been scaled assuming that a S/N=10 corresponds to 28 mag in these bands.} We considered only a signal-to-noise of 10 to limit the effects of redshift uncertainties and bad photometry. Indeed, in B16 the number of outliers in redshift (galaxies with $|z_{phot}-z_{fiducial}|/(1+z_{fiducial})>$0.15) for this S/N value was less than 1$\%$ at all considered redshifts. Therefore, the variations in the galaxy property estimations obtained here are mainly due to degeneracies between galaxy templates and they are not expected to be solved just with a higher integration time. Differences in the S/N value are expected not to change significantly the results of this work, as long as $S/N\gtrsim10$. \par
A 28 AB mag with a S$/N=$10 in MIRI corresponds to $\sim$196 hours per pointing of exposure time with F560W and a low-level backgrounds.
However, as explained in more detail in section \ref{sec:mass_StN5}, considering a S/N$=$5 for a 28 AB magnitude on the MIRI bands will not create a major difference on the stellar mass estimation and it will be possible with $\sim$47 hours per pointing of exposure time. \par
Each mock flux is randomized 100 times within the error bars and we considered as non-detection every flux below the 2$\sigma$ level. As a conservative approach, when running the SED-fitting code, we adopted a 3$\sigma$ upper limit, rather than 2$\sigma$, in all cases of non-detections.\par
More details about the \emph{JWST} photometry and redshift derivation are presented in B16.\par

\subsection{Galaxy properties derivation}
We used the expected \emph{JWST} photometry for our simulated galaxy sample to test how accurately it is possible to derive different properties, i.e. stellar mass, age, color excess and sSFR. As in B16, we performed these tests with different combinations of \emph{JWST} bands:
\begin{itemize}
\item 8 NIRCam broad bands
\item 8 NIRCam broad bands and 2 MIRI bands (F560W and F770W)
\item 8 NIRCam broad bands and MIRI F560W only
\item 8 NIRCam broad bands and MIRI F770W only
\end{itemize} 
We considered these band combinations to understand in which situations NIRCam broad-band filters are sufficient to properly recover galaxy properties and in which cases, adding the two shortest-wavelength MIRI filters, can significantly improve the estimations. We considered only these two MIRI broad-band filters, because they are the most sensitive ones and, therefore, they are those which will be mostly used in high-z galaxy surveys. As deep observations with MIRI are significantly more time consuming that with NIRCam, because of different detector technology in the near and mid-IR, observation strategies may opt to use only one of the two bands. Therefore, we considered this situation by analyzing also filter combinations with only one of the two MIRI filters.  \par
For all filter combinations, we ran the public code \textit{LePhare} \citep{Arnouts1999,Ilbert2006} to derive stellar mass, age, color excess and sSFR allowing for a large variety of templates.  A full list of parameter values used for the SED fitting is presented in table \ref{tab:param_out}. \par
First, when running \textit{LePhare}, we considered a library containing the same templates used to derive \emph{JWST} photometry, i.e. BC03 templates with incorporated emission lines and the \textit{Yggdrasil} templates, which a more refined grid of color excess values and z=0-11.  However, in BC03 templates, we limited the used templates to only those corresponding to the pre-determined photometric redshift. For these templates, we added manually the main emission lines, whose rest-frame equivalent widths change with redshift and metallicity (see B16 for more details). However, we note that the use of templates that change with redshifts introduce an unnecessary level of degeneracy in parameter space. Albeit the redshifts are correctly recovered in the vast majority of cases, assigning the ``wrong" line equivalent widths could create more scatter in the derived stellar mass. In particular, this happens when it is not possible to determine the level of the continuum and, therefore, the equivalent width of the emission lines. This is the case when there is not a clean measurement of the near-IR continuum at $\lambda>4000\rm\AA$, without contamination by emission lines, e.g. when using only 8 NIRCam broad bands at z=7. In this case, the level of the continuum can be determined observing at rest-frame $\lambda>4000\rm\AA$ (i.e. with the F770W band at z=7), but it is necessary to increase the number of used filters to cover all possibilities in a wide range of redshifts. However, since the redshift estimation is mostly not affected by this problem and it is already possible to derive a good photometric redshift using 8 NIRCam bands if the S/N$\geqslant$10, we decided to first estimate the photometric redshift using all templates and then derive the stellar mass, together with age and color excess, limiting the templates to the ones corresponding to the derived redshift. \par
In the SED fitting, we included also templates of older galaxies and higher extinction values than those expected at z$>$7, in order to allow for degeneracies between redshift, dust and age to be manifested. In particular, we included templates from BC03 with solar metallicity, ages from 0.01 to 5~Gyr, exponentially declining star formation histories with different characteristic times $\tau$ from 0.01 to 10~Gyr and we applied color excess following the Calzetti et al. reddening law \citep{Calzetti2000} with extinction values E(B-V)$=0-1$, with a step of 0.05 mag. \par
Each galaxy in the sample has 100 sets of photometry, obtained randomizing fluxes within each own error bar. We derived the median of the photometric redshifts of the 100 runs of each galaxy and we recovered the parameter values of the median z$_{phot}$ solutions, despite its equivalence with the fiducial redshift, albeit the redshifts themselves are correctly recovered in most cases. So, we proceeded in two steps: first, we derived the median stellar mass among the recovered solutions, then we retrieved the parameters values of the median stellar mass solution. To summarize, we considered the parameters set of the template corresponding to the median photometric redshift and median stellar mass, as the final output values of stellar mass, color excess, age and sSFR of each galaxy. While we considered as fiducial values the input values of each used SED model. \par

\begin{deluxetable*}{cccc}
\tablecaption{Parameter values used for the SED fitting runs of the simulated galaxies at z$=7-10$ that are analyzed in this work. For both types of templates we used the original template, BC03 or \emph{Yggdrasil}, with a denser grid of color excess values and a wider redshift range than the original templates used to derived the mock photometry (see table \ref{tab:param}). Moreover, in both cases, we considered also an additional BC03 templates with old ages. \label{tab:param_out}}
\tablecolumns{4}
\tablewidth{0pt}
\tablehead{
\colhead{Parameter} &
\colhead{Values (BC03)} &
\colhead{Values (Ygg)} &
\colhead{Values (Additional BC03)}
}
\startdata
metallicity & Z$_{\odot}$, 0.4Z$_{\odot}$, 0.2Z$_{\odot}$,0.02Z$_{\odot}$ & Z$_{\odot}$,0.4Z$_{\odot}$,0.2Z$_{\odot}$,0.02Z$_{\odot}$  & Z$_{\odot}$,0.4Z$_{\odot}$,0.2Z$_{\odot}$,0.02Z$_{\odot}$\\
SFH type & declining & step function & declining\\
SFH [Gyr]& 0.01,0.1,1,10 & 0.01,0.03,0.1 & 0.01,0.1,1,10\\
$f_{\rm cov}$ & -- & 0,1 & --\\
E(B-V)\tablenotemark{a} & 0, 0.05, 0.1,...,1  & 0, 0.05, 0.1,...,1 & 0, 0.05, 0.1,...,1\\
age [Gyr] & 0.01,0.05,0.2,0.4,0.6 & 0.01,0.05,0.2,0.4,0.6 & 1,1.5,2.5,5\\
$z$\tablenotemark{b} & 0-11 & 0-11 & 0-11\\
\enddata
\tablenotetext{a}{following the Calzetti et al. reddening law \citep{Calzetti2000}}
\tablenotetext{b}{this is the redshift range in the SED fitting run}
\end{deluxetable*}

\section{Results}\label{sec:results}
\subsection{Stellar masses}
In this section we present our stellar mass results, derived using NIRCam broad bands alone and adding alternatively the F560W and F770W MIRI bands. All masses are derived from the normalization of each SED template, using all available bands. It is important to remember that, at S/N$=$10, the number of outliers in redshift ($|z_{output}-z_{fiducial}|/(1+z_{fiducial})>$0.15) is less than 1$\%$ for both BC03 and \emph{Yggdrasil} templates with all band combinations \citep{Bisigello2016}.
Because the considered redshifts are four fixed values ($z=7$, 8, 9 and 10), we showed  the output mass estimation for each case separately and for each plot we quote the mean and the r.m.s values of the $\Delta(log(M^{*}))=log(M^{*}_{output})-log(M^{*}_{fiducial})$ distribution, where fiducial values for the stellar mass are the input values of each template. Our results are shown in Fig. \ref{fig:mass_BC03}-\ref{fig:mass_Ygg_EL} for BC03 and \emph{Yggdrasil} templates separately. 

\subsubsection{Galaxies simulated with BC03 templates}\label{sec:mass_BC03}
\begin{figure*}[ht!]
\center{
\includegraphics[width=1\linewidth, keepaspectratio]{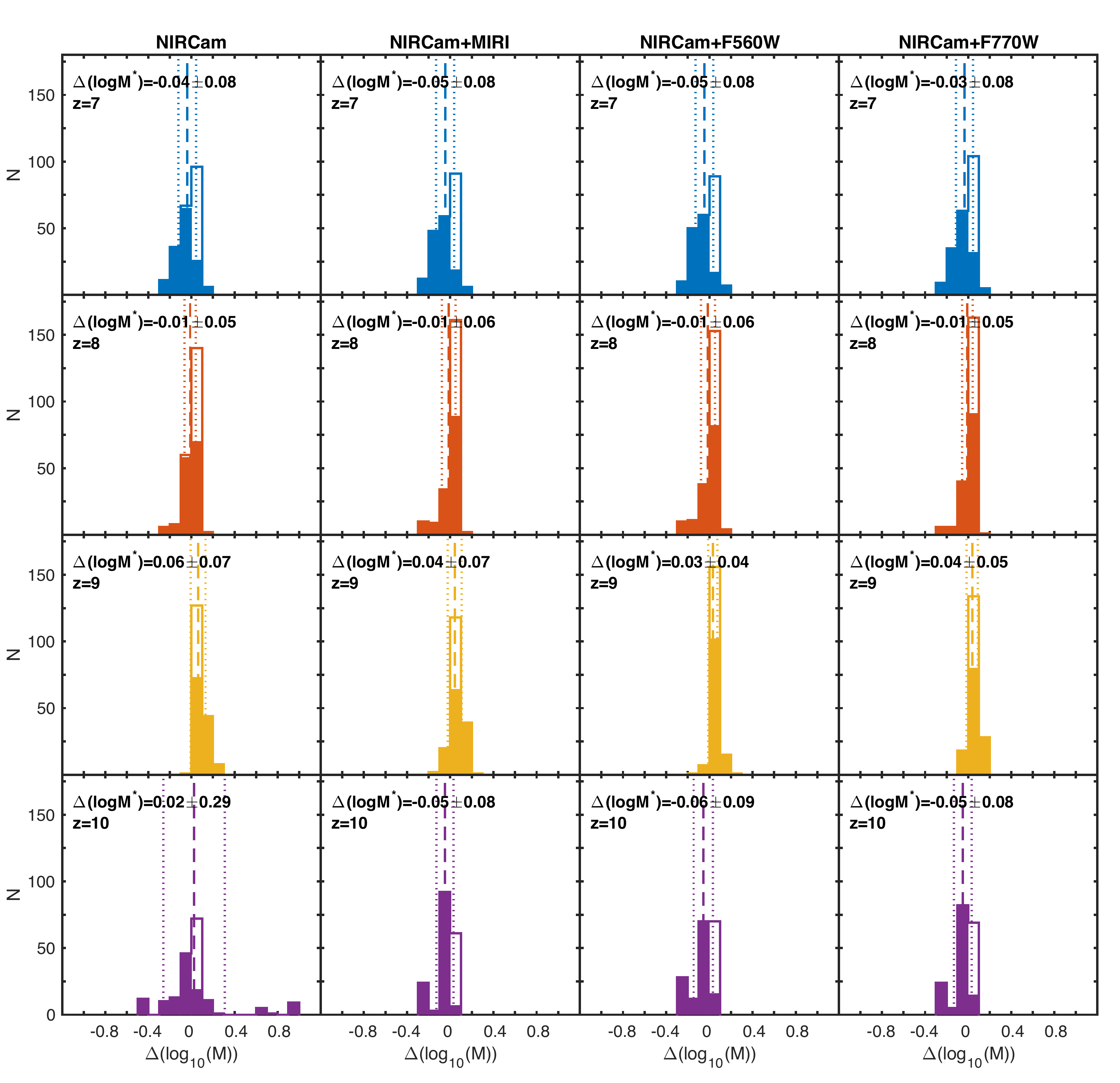}}
\caption{ Differences between the derived and the fiducial stellar mass for the BC03 simulated galaxies at different fixed redshifts, i.e. $\Delta log_{10}(M^{*})=log_{10}(M^{*}_{output})-log_{10}(M^{*}_{fiducial})$. \textit{From top to bottom}: redshifts $z=7, 8, 9$ and 10.  Stellar masses in each column are obtained with different combinations of \emph{JWST} bands. \textit{From left to right:} 8 NIRCam broad bands; 8 NIRCam broad bands, MIRI F560W and MIRI F770W; 8 NIRCam broad bands and MIRI F560W only; 8 NIRCam broad bands and MIRI F770W only. The vertical lines indicate the mean and the 1 $\sigma_{logM^{*}}$ values, which are quoted at the top-left of each panel. The colored histograms represent galaxies with emission lines, while the white ones represent galaxies without emission lines. \label{fig:mass_BC03}}
\end{figure*}

For the BC03 templates, the results for the stellar mass recovery are shown in Fig. \ref{fig:mass_BC03}. The outlier fractions for each \emph{JWST} filter combination are listed in Table \ref{tab:M_BC03}. Outliers are defined as galaxies with $\Delta log(M)=log_{10}(M^{*}_{output})-log_{10}(M^{*}_{fiducial})>$3$\sigma_{logM^{*}}$, with $\sigma_{logM^{*}}=$0.04 dex, which is the minimum $\sigma_{logM^{*}}$ obtained with all considered filter combinations.\par
First, we analyzed the results considering only 8 NIRCam broad-band filters. The mass difference distribution is quite narrow up to $z=9$, with $\sigma_{logM^{*}}=$0.05-0.08 dex, but it becomes broad at the highest redshift, with $\sigma_{logM^{*}}=$0.29 dex. The fraction of outliers is high at all redshifts, at $z=7-9$ it is between 5.6$\%$ and 25.6$\%$ and it is even higher (31.1$\%$) at $z=10$. The wings of the distribution are almost totally composed of galaxies with emission lines, that also made almost all the outliers. We will analyze galaxies with emission lines further in Section \ref{sec:mass_Ygg}.  \par
Second, we analyzed the results derived by using both MIRI bands together with the 8 NIRCam broad-band filters. At z$=7-8$, the incorporation of the MIRI bands is not improving the already small r.m.s. values obtained with NIRCam alone. The number of outliers when adding MIRI bands is slightly higher than when considering only NIRCam bands, however the stellar mass offset is less then 0.15 dex for the majority of cases, and never over 0.3 dex. Moreover, when the stellar mass estimation worsen adding the two MIRI bands, the estimation of other parameters, such as age or color excess, generally improves (see Section \ref{sec:age_BC3}).
At $z=9$, instead, the r.m.s. value does not significantly change, but the mean does decrease as well as the number of outliers, which becomes $\sim$16$\%$. \par
At $z=10$, the importance of the MIRI bands becomes more evident. Indeed, the r.m.s. value becomes similar to other redshifts, decreasing from 0.29 dex with 8 NIRCam bands to 0.08 dex when adding the two MIRI bands. Also the fractions of outliers decrease from 31.1$\%$ with 8 NIRCam bands to 13.3$\%$ with the two MIRI bands. This happens because few NIRCam bands cover a rest-frame wavelength redward of the 4000 $\rm\AA$ break at increasing redshifts, i.e. at $z=10$ no NIRCam broad-bands purely cover $\lambda>$4000$\rm\AA$. \par
When the MIRI bands are considered one at a time, the r.m.s. values are generally similar to the case with both MIRI bands at all redshifts as well as the fraction of outliers, with small differences in one or the other band depending on the redshift. \par
Overall, when considering BC03 templates and NIRCam observations alone, the stellar mass estimation is generally good for galaxies without emission lines and for galaxies with emission lines at $z\leq9$, while it becomes more difficult at $z=10$. Indeed, at $z=10$ no NIRCam band purely cover rest-frame $\lambda>4000\rm\AA$ and it is difficult to set the level of the continuum and , therefore, derive the stellar mass. Adding the two MIRI bands does not significantly improve the already good mass estimation at the lowest redshifts, while it improves it substantially at $z=10$ by adding more information at $\lambda>4000\rm\AA$.

\begin{deluxetable*}{ccccc}
\tablecaption{ Number (percentage) of outliers in stellar mass among the 792 (216 at each $z=7$ and 8 and 180 at each $z=9$ and 10) galaxies simulated with BC03 templates and manual addition of emission lines, for different combination of \emph{JWST} filter combinations and different redshifts. Between square brackets there are the number of galaxies with emission lines among the outliers. Outliers are defined as those values beyond 3$\sigma_{logM^{*}}$ of the distribution $\Delta log_{10}(M^{*})=log_{10}(M^{*}_{output})-log_{10}(M^{*}_{fiducial})$, with $\sigma_{logM^{*}}=$0.04 dex.\label{tab:M_BC03}}
\tablecolumns{4}
\tablewidth{0pt}
\tablehead{
\colhead{Bands} &
 \colhead{N$_{outlier,z=7}$} &
 \colhead{N$_{outlier,z=8}$} &
 \colhead{N$_{outlier,z=9}$} &
 \colhead{N$_{outlier,z=10}$} 
}
\startdata
8 NIRCam broad bands &  41 [41] (19.0$\%$) & 12 [12] (5.6$\%$) & 46 [46] (25.6$\%$) & 56 [55] (31.1$\%$)\\
8 NIRCam bands+MIRI F560W, F770W &  50 [50] (23.1$\%$) & 21 [21] (9.7$\%$) & 28 [28] (15.6$\%$) & 24 [24] (13.3$\%$)\\
8 NIRCam bands+MIRI F560W & 50 [50] (23.1$\%$) & 24 [24] (11.1$\%$) & 10 [10] (5.6$\%$) & 34 [34] (18.9$\%$)\\
8 NIRCam bands+MIRI F770W  & 39 [39] (18.1$\%$) & 11 [11] (5.1$\%$) & 14 [14] (7.8$\%$) & 25 [25] (13.9$\%$)\\
\enddata
\end{deluxetable*}

\subsubsection{Galaxies simulated with Yggdrasil templates}\label{sec:mass_Ygg}
\begin{figure*}[ht!]
\center{
\includegraphics[width=1\linewidth, keepaspectratio]{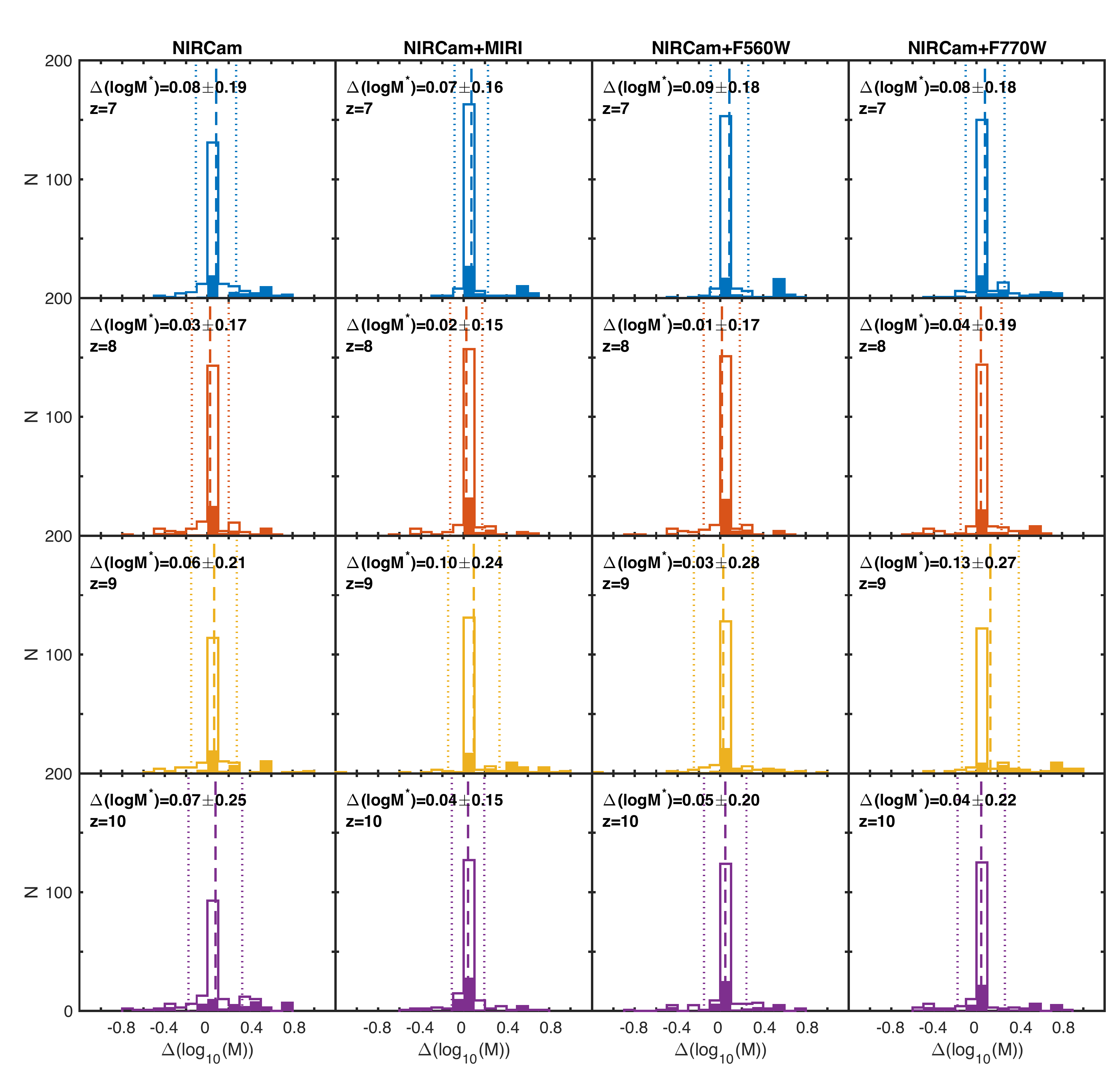}}
\caption{ Differences between the derived and the fiducial stellar mass for the \emph{Yggdrasil} simulated galaxies at different fixed redshifts, i.e. $\Delta log_{10}(M^{*})=log_{10}(M^{*}_{output})-log_{10}(M^{*}_{fiducial})$.  \textit{From top to bottom}: redshifts $z=7, 8, 9$ and 10.  Stellar masses in each column are obtained with different combinations of \emph{JWST} bands. \textit{From left to right:} 8 NIRCam broad bands; 8 NIRCam broad bands, MIRI F560W and MIRI F770W; 8 NIRCam broad bands and MIRI F560W only; 8 NIRCam broad bands and MIRI F770W only. The vertical lines indicate the mean and the 1 $\sigma_{logM^{*}}$ values, which are quoted at the top-left of each panel. The colored histograms represent galaxies with emission lines, while the white ones represent galaxies without emission lines. \label{fig:mass_Ygg}}
\end{figure*}

Figure \ref{fig:mass_Ygg} shows the distribution of the difference between the recovered stellar masses and the fiducial ones for galaxies simulated with the \textit{Yggdrasil} templates at each fixed redshift and for all \emph{JWST} filter combinations. The fractions of outliers of the full sample of \emph{Yggdrasil} simulated galaxies are listed in table \ref{tab:M_Ygg}. Outliers are defined as galaxies with $\Delta log(M)=log(M^{*}_{output})-log(M^{*}_{fiducial})>3\sigma_{logM^{*}}$, with $\sigma_{logM^{*}}=$0.15 dex, which is the minimum $\sigma_{logM^{*}}$ obtained with all considered filter combinations. \par
When considering only 8 NIRCam bands, the $\Delta$log(M) distributions appear peaky with long tails, particularly towards positive values. The r.m.s. values are similar at $z=7$ and 8, namely $\sigma_{logM^{*}}=$0.19 and 0.17 dex respectively, and they increase at higher redshifts up to 0.25 dex at $z=10$. The fraction of outliers ranges between 4.5$\%$ at $z=8$ and 12.6$\%$ at $z=10$.  \par
When adding both MIRI bands, the mass estimation slightly improves at all redshifts, but at $z=9$. In particular, the main improvement is in the number of outliers at z$ =$10 which is reduced by more than a half. Adding the two MIRI bands at z=9 does not change significantly the mass estimation, except for galaxies with emission lines for which the estimation became worse. However, the estimation of other parameters, such as age and color excess, generally improves when the stellar mass does not (see \ref{sec:age_Ygg}). Indeed, for this particular redshift and for galaxies with emission lines, the stellar mass estimation, as well as other parameters, is particularly difficult also when including the MIRI bands. This is because emission lines are present in all bands at $\lambda>4000\rm\AA$ and, therefore, it is not possible to set the level of the continuum. \par
When analyzing the two MIRI bands separately, we see that they have similar roles, but F560W decreases the r.m.s. and the fraction of outliers slightly more than F770W at $z>7$. \par
\begin{figure*}[ht!]
\center{
\includegraphics[width=1\linewidth, keepaspectratio]{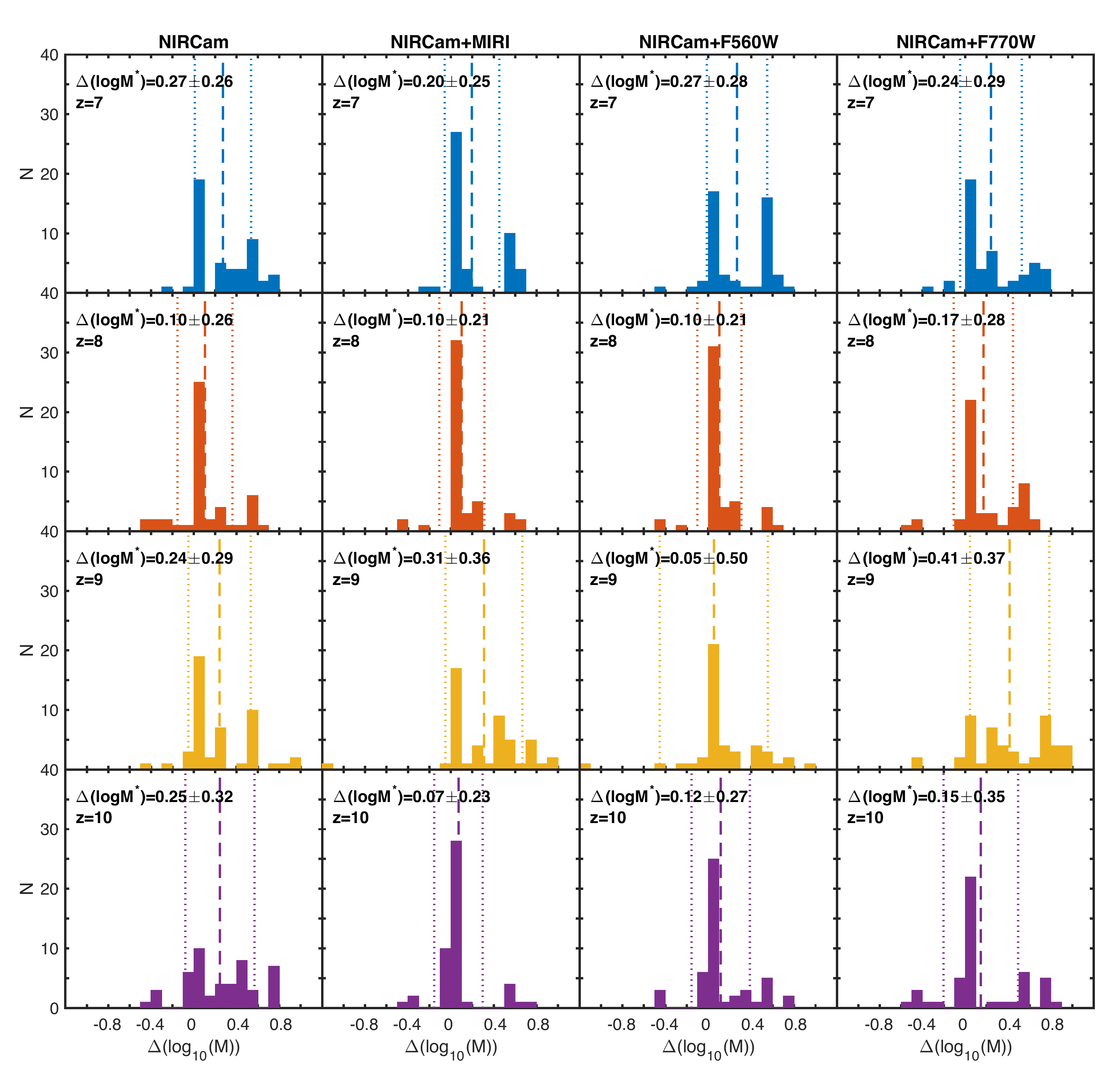}}
\caption{ Zoom in of Fig. \ref{fig:mass_Ygg} showing only galaxies with emission lines. \label{fig:mass_Ygg_EL}}
\end{figure*}

Figure \ref{fig:mass_Ygg_EL} shows the distribution of the difference between the recovered and the fiducial stellar mass, but only for galaxies with emission lines, which are young galaxies with covering factor f$_{cov}>$0. The number of outliers is defined as galaxies with $log(M^{*}_{output})-log(M^{*}_{fiducial})>3\sigma_{logM^{*}}$, with $\sigma_{logM^{*}}=$0.15 dex, similarly to the full \emph{Yggdrasil} sample, and they are reported in table \ref{tab:M_Ygg} between square bracket. \par
When considering only 8 NIRCam bands, the stellar mass is generally overestimated, with mean values between 0.10 dex and 0.27 dex. The r.m.s. values are higher than the values of the general sample, ranging between 0.26 dex and 0.32 dex, with a minimum value at $z\leq8$. 
Among all outliers, galaxies with emission lines are the majority, but also a small fraction of galaxies without emission lines are present, differently to BC03 templates. Among the outliers with no emission lines, the majority (22 out of 31) are red and quenched galaxies and the remaining are blue SF galaxies with f$_{cov}=$0 that are not present among BC03 templates. \par
Adding the MIRI bands produces small improvements on the stellar mass recovery at z$=$7-8, while, as for the general sample, the two MIRI bands have the largest impact at z$=$10, with the r.m.s. that decreases of 0.09 dex while the mean changes from 0.25 dex to 0.07 dex. Similarly, when looking at the number of outliers, at $z<9$ adding the two MIRI bands creates small differences ($\sim20\%$) while at $z=10$ the number of outliers is reduced to almost a half. At $z=9$ the number of outliers with emission lines increases.\par
When considering the two MIRI bands separately, F560W improves the mass estimation slightly more than the other band, with smaller r.m.s. values at all redshifts, but at z$=9$, and lower number of outliers at $z>7$. \par
Overall, when considering \emph{Yggdrasil} templates, the stellar mass estimation derived with 8 NIRCam bands is generally good at z$=7-9$ but it worsens with increasing redshift. Moreover, it is particularly difficult when galaxies have nebular emission lines. MIRI bands slightly improve the mass estimation of the general sample and it particularly improves it at $z=10$, because MIRI bands cover the rest-frame $\lambda>4000\rm\AA$ which, otherwise, is purely covered by no NIRCam bands at this redshift. Moreover, adding MIRI bands improve the stellar mass recovery of galaxies with emission lines at all redshifts, except at $z=9$.

\begin{deluxetable*}{ccccc}
\tablecaption{ Number (percentage) of outliers in stellar mass among the 750 (201 at each $z=7$ and 8 and 174 at each $z=9$ and 10) galaxies simulated with \textit{Yggdrasil} templates, 192 of which are with emission lines, for different combination of \emph{JWST} filter combinations and different redshifts. Between square brackets there are the number of galaxies with emission lines among the outliers. Outliers are defined as those values beyond 3$\sigma_{logM^{*}}$ of the distribution $\Delta log_{10}(M^{*})=log_{10}(M^{*}_{output})-log_{10}(M^{*}_{fiducial})$, with $\sigma_{log_{10}M^{*}}=$0.15 dex.\label{tab:M_Ygg}}
\tablecolumns{4}
\tablewidth{0pt}
\tablehead{
\colhead{Bands} &
 \colhead{N$_{outlier,z=7}$} &
 \colhead{N$_{outlier,z=8}$} &
 \colhead{N$_{outlier,z=9}$} &
 \colhead{N$_{outlier,z=10}$} 
}
\startdata
8 NIRCam broad bands &  16 [15] (8.0$\%$) & 9 [7] (4.5$\%$) & 16 [15] (9.2$\%$) & 22 [16] (12.6$\%$)\\
8 NIRCam bands+MIRI F560W, F770W &  16 [14] (8.0$\%$) & 6 [5] (3.0$\%$) & 21 [20] (12.1$\%$) & 9 [7] (5.2$\%$)\\
8 NIRCam bands+MIRI F560W & 22 [22] (10.9$\%$) & 9 [6] (4.5$\%$) & 13 [13] (7.5$\%$) & 14 [10] (8.0$\%$)\\
8 NIRCam bands+MIRI F770W  & 15 [14] (7.5$\%$) & 15 [13] (7.5$\%$) & 22 [22] (12.6$\%$) & 20 [15] (11.5$\%$)\\
\enddata
\end{deluxetable*}

\subsubsection{Stellar mass recovery with S/N$=$5 with MIRI}\label{sec:mass_StN5}
As explained before, the integration time for reaching a S/N$=$10 at 28 AB mag with MIRI are significantly longer that for NIRCam. However, when considering a S$/N=$5, instead of a S$/N=$10, at 28 AB mag for the MIRI bands and leaving a S/N$=$10 with all NIRCam bands, the differences in the derived stellar masses are $<5\%$ for 94$\%$ of the sample of \textit{Yggdrasil} templates. This percentage varies a little with redshift from $\sim$96$\%$ at z$=$7 to $\sim$92$\%$ at z$=$10. For BC03 templates the differences in the derived stellar masses are $<5\%$ for 79$\%$ of the sample at z$=$7, 87$\%$ at z$=$8, 56$\%$ at z$=$9 and 85$\%$ at z$=$10. \par
Reaching mag$=$28 with a S/N$=$5 with MIRI is indeed possible, i.e. $\sim$38 ($\sim$63) hours of exposure time with F560W band and a low-level (medium-level) background, and the stellar mass derived is similar to the one obtained considering a S$/N=$10 with MIRI for the majority of cases.

\subsection{Age and color excess}\label{sec:age}
In this section we present our results of age and color excess recovery for galaxies derived both with BC03 templates and with \emph{Yggdrasil} models. Because of the age-dust degeneracy, we decided to analyze the two properties together. Figures \ref{fig:ageBC03} - \ref{fig:ageYgg_EL} show the difference between the derived and fiducial ages compared to the difference between the derived and fiducial color excess values at each fixed redshift, for BC03 and \textit{Yggdrasil} templates. For each plot we quote the mean and r.m.s. values of both the age difference ($t_{output}-t_{fiducial}$) and the color excess difference ($E(B-V)_{output}-E(B-V)_{fiducial}$) distributions. As for the stellar masses, we presented our results derived using observations with only 8 NIRCam broad-band filters, adding both MIRI bands and only one of the two separately.

\subsubsection{Galaxies simulated with BC03 templates}\label{sec:age_BC3}
\begin{figure*}[ht!]
\center{
\includegraphics[width=1\linewidth, keepaspectratio]{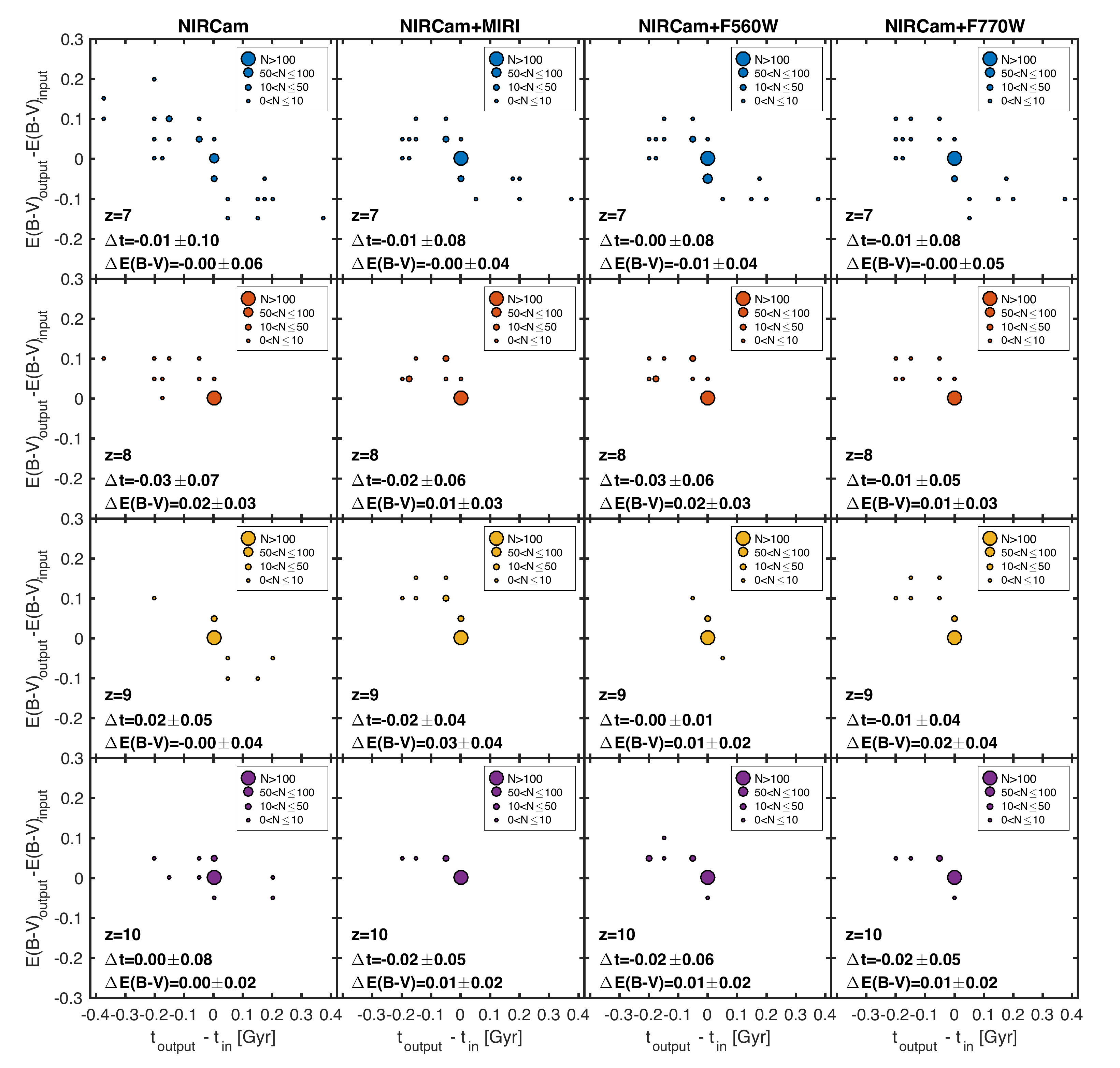}}
\caption{ Difference between the derived age and the fiducial one, $t_{output}-t_{fiducial}$, against the difference between the derived color excess and the fiducial one,  E(B-V)$_{output}$-E(B-V)$_{fiducial}$, for galaxies simulated with BC03 templates at different fixed redshifts. \textit{From top to bottom}: redshifts $z=7, 8, 9$ and 10.  Ages in each column are obtained with different combinations of bands. \textit{From left to right:} 8 NIRCam broad bands; 8 NIRCam broad bands, MIRI F560W and MIRI F770W; 8 NIRCam broad bands and MIRI F560W only; 8 NIRCam broad bands and MIRI F770W only.  
Galaxies are divided into bins of 25 Myr in age and 0.025 mag in color excess and the size of each circle represent the number of galaxies inside each bin. The size of the dots increase with the number of galaxies inside the respective bin. Mean and r.m.s. values of the age and the color excess estimation are shown at the bottom-left of each panel. \label{fig:ageBC03}}
\end{figure*}

In Figure \ref{fig:ageBC03} we show the comparison between fiducial and output values, both for age and color excess, for BC03 templates. Outliers are defined as galaxies with $|t_{output}-t_{fiducial}|>3 \sigma_{t}$ ($|E(B-V)_{output}-E(B-V)_{fiducial}|>3 \sigma_{E(B-V)}$), with $ \sigma_{t}=$0.01 Gyr ($ \sigma_{E(B-V)}=$0.02 mag) which is the minimum r.m.s. value for the age (color excess) obtained with all considered filter combinations.\par
When considering only 8 NIRCam bands, the color excess is estimated within $|\Delta E(B-V)|<$0.2 mag for all objects at all redshifts, the mean of the distribution is $<$0.02 mag and the r.m.s. values are 0.02-0.06 mag, with the highest value at $z=7$. The number of outliers is high at z=7 and it decreases with redshifts, from $25.5\%$ to $0.0\%$. On the other hand, the age difference distribution has r.m.s. values between 0.05 (at z=9) and 0.10 (at z=7) and the mean values are between  -0.03 Gyr and 0.02 Gyr. However, long tails are present up to $|\Delta t|=$0.4 Gyr and the number of outliers ranges from a minimum of 15.0$\%$ at z=9 and a maximum of 36.1$\%$ at z=7. In general, the age and the color excess r.m.s. values decrease with the redshift also because the age-dust degeneracy is reduced due to the decreasing age of the Universe. \par
When adding both MIRI bands, the r.m.s. of the color excess difference does not change significantly, but the numbers of outliers decrease at $z\leq8$. On the other hand, the age r.m.s. slightly decreases at all redshifts, but, the number of outliers decreases only at $z=7$ and 10.  Moreover, at $z>7$ the age is never overestimated while the color excess is never underestimated. \par
The two MIRI bands play similar roles when considered separately and none of them improves the age or the color excess estimation particularly more than the other one at all redshifts. \par
Similarly to the mass estimation, most of the outliers, both in age and color excess, are galaxies with emission lines.\par
To conclude, with only NIRCam bands, both the color excess and age estimation distribution present long tails and extreme outliers. The two MIRI bands improve the age estimations, reducing mainly the r.m.s., while they mainly reduce the number of outliers in the color excess estimation. \par

\subsubsection{Galaxies simulated with \textit{Yggdrasil} templates}\label{sec:age_Ygg}
\begin{figure*}[ht!]
\center{
\includegraphics[width=1\linewidth, keepaspectratio]{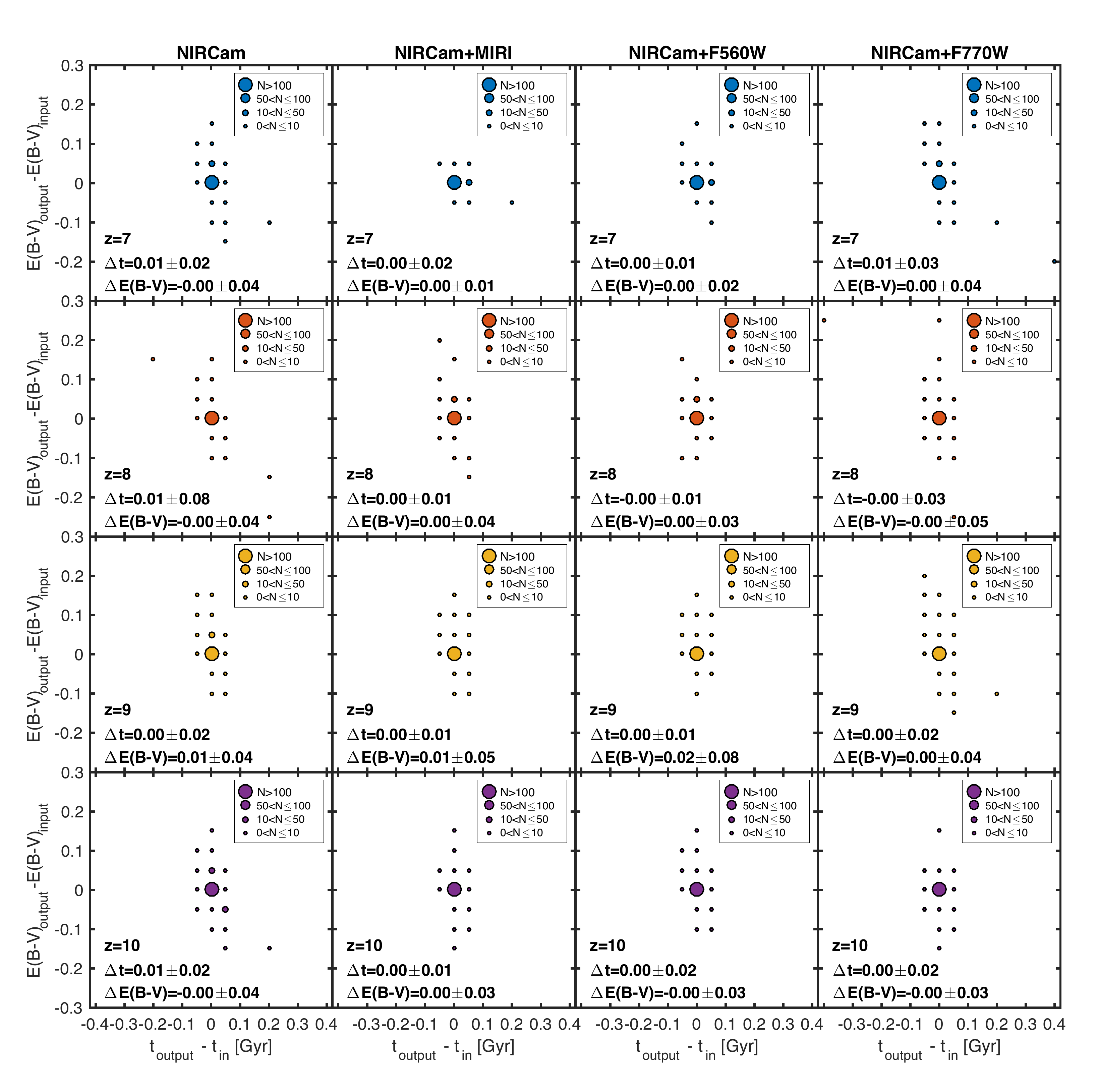}}
\caption{ Difference between the derived age and the fiducial one, $t_{output}-t_{fiducial}$, against the difference between the derived color excess and the fiducial one,  E(B-V)$_{output}$-E(B-V)$_{fiducial}$, for galaxies simulated with \textit{Yggdrasil} templates at different fixed redshifts. \textit{From top to bottom}: redshifts $z=7, 8, 9$ and 10.  Ages in each column are obtained with different combinations of bands. \textit{From left to right:} 8 NIRCam broad bands; 8 NIRCam broad bands, MIRI F560W and MIRI F770W; 8 NIRCam broad bands and MIRI F560W only; 8 NIRCam broad bands and MIRI F770W only. Galaxies are divided into bins of 25 Myr in age and 0.025 mag in color excess and the size of each circle represent the number of galaxies inside each bin. The size of the dots increase with the number of galaxies inside the respective bin. Mean and r.m.s. values of the age and the color excess estimation are shown at the bottom-left of each panel. \label{fig:ageYgg}}
\end{figure*}

In Figure \ref{fig:ageYgg} we show the comparison between fiducial and output values, both for age and color excess, for \textit{Yggdrasil} templates, for each redshift and different band combinations.  Outliers are defined as galaxies with $|t_{output}-t_{fiducial}|<3 \sigma_{t}$ ($|E(B-V)_{output}-E(B-V)_{fiducial}|<3 \sigma_{E(B-V)}$), with $ \sigma_{t}=$0.01 Gyr ($ \sigma_{E(B-V)}=$0.01 mag) which is the minimum r.m.s. value obtained for the age (color excess) with all considered filter combinations.\par
When considering only 8 NIRCam bands, the color excess estimation is similar at all redshifts, with $\sigma_{E(B-V)}=$0.04 mag. The number of outliers is also similar at all redshifts ($\sim20-30\%$), with a minimum value at $z=8$. The age difference distribution has $\sigma_{t}=0.02-0.08$ Gyr, with a maximum value at z$=8$, while the number of outliers is between the 12.9$\%$ and the 25.3$\%$, increasing with redshift. \par
When including the two MIRI bands, both the age and the color excess estimation are generally improved.
In particular, for the color excess, the r.m.s. value greatly decreases at $z=7$ and it does not change significantly at the other redshifts, while the number of outliers decreases at all redshifts and in particular at $z=7$ and 10, where it is reduced to less than half. At $z=9$, the r.m.s. slightly increases, because of a single galaxy for which the color excess is highly overestimated (0.04 mag), otherwise, without this outlier, it would remain unchanged respect to the value derive with only NIRCam broad-bands.  When considering the age estimation, the r.m.s. values became 0.01 Gyr at all redshifts, decreasing the values derived with only 8 NIRCam bands, except at $z=7$ where it remain 0.2 Gyr. The number of outliers decreases from 13-25$\%$, when considering only NIRCam bands, to $\sim10\%$, with both MIRI bands. Moreover, there are no galaxies with $|\Delta t|>$0.1 Gyr at z$>$8. \par
The two MIRI bands improves age and color excess similarly, but the F560W decreases the r.m.s. of both the age difference and the color excess difference a bit more than the other band. \par
\begin{figure*}[ht!]
\center{
\includegraphics[width=1\linewidth, keepaspectratio]{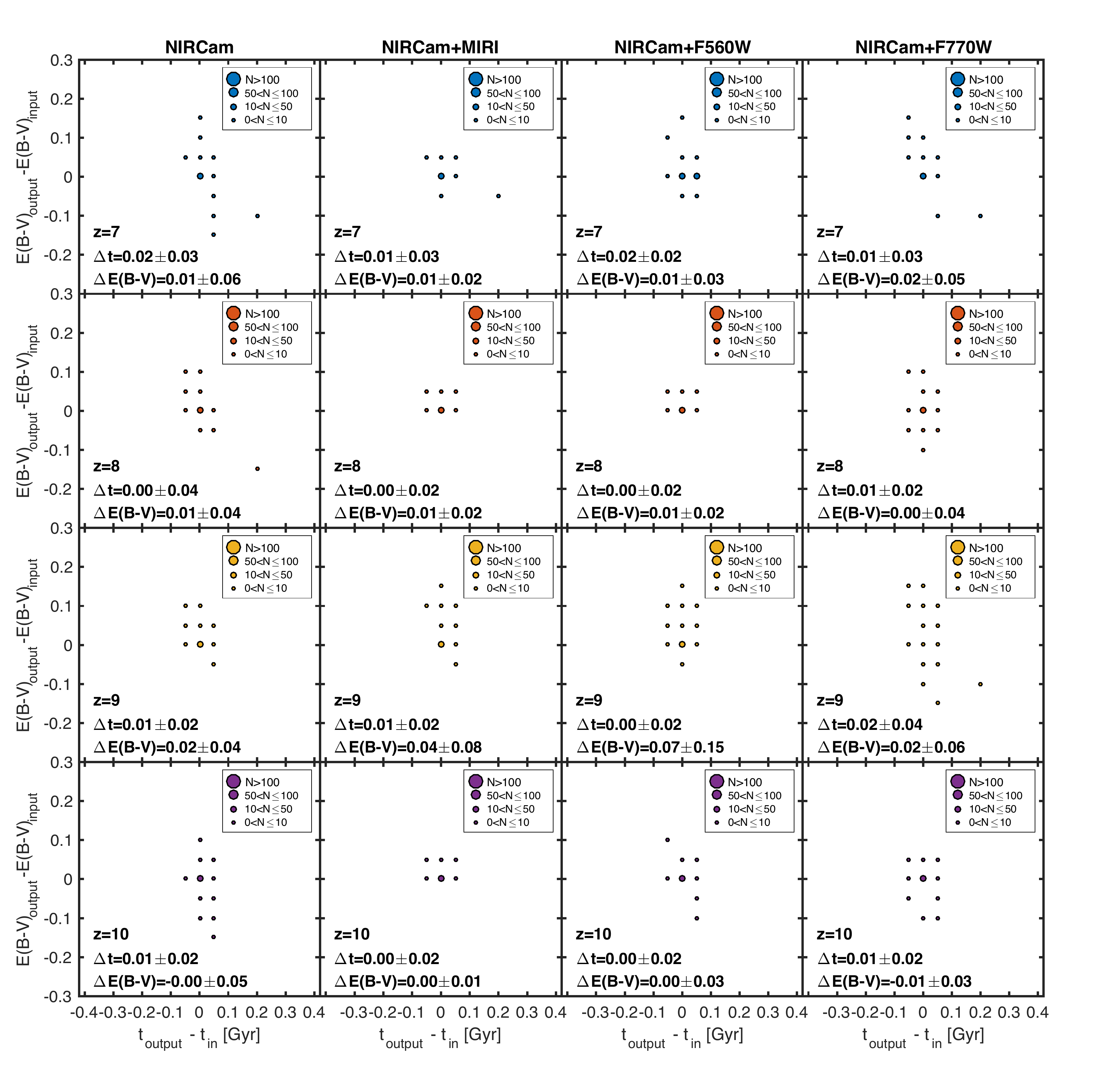}}
\caption{ Difference between the derived age and the fiducial one, $t_{output}-t_{fiducial}$, against the difference between the derived color excess and the fiducial one,  E(B-V)$_{output}$-E(B-V)$_{fiducial}$, for galaxies simulated with \textit{Yggdrasil} templates and emission lines at different fixed redshifts. \textit{From top to bottom}: redshifts $z=7, 8, 9$ and 10.  Ages in each column are obtained with different combinations of bands. \textit{From left to right:} 8 NIRCam broad bands; 8 NIRCam broad bands, MIRI F560W and MIRI F770W; 8 NIRCam broad bands and MIRI F560W only; 8 NIRCam broad bands and MIRI F770W only. Galaxies are divided into bins of 25 Myr in age and 0.025 mag in color excess and the size of each circle represent the number of galaxies inside each bin. The size of the dots increase with the number of galaxies inside the respective bin. Mean and r.m.s. values of the age and the color excess estimation are shown at the bottom-left of each panel. \label{fig:ageYgg_EL}}
\end{figure*}

In Figure \ref{fig:ageYgg_EL},  we show the comparison between fiducial and output values, both for age and color excess, but only for galaxies with emission lines, which are young star-forming galaxies with covering factor f$_{cov}>0$. \par
The color excess estimation for galaxies with emission lines does not present significant differences to the full \emph{Yggdrasil} sample, both when considering only 8 NIRCam bands and when adding the two MIRI bands. The same happens generally to the age estimation. As for the stellar mass, not all outliers in age and color excess are galaxies with emission lines.
\par
To conclude, when considering only the 8 NIRCam bands, both the age and the color excess difference distributions present long tails. Adding the two MIRI bands  improves both the age and the color excess estimation, slightly decreasing the r.m.s. values, but mainly reducing the number of outliers.  \par

\subsection{Specific star formation rates}\label{sec:sSFR}
In this section we present our results for the sSFR of star-forming galaxies derived both with BC03 templates and with \textit{Yggdrasil} models. As it is explained later in more details, in BC03 and \textit{Yggdrasil} SED templates SFR$\propto$M$^{*}$. So, to decouple stellar mass effects, we considered the sSFR instead of the SFR.  Results are shown in figures \ref{fig:sSFR_BC03}-\ref{fig:sSFR_Ygg} at the four fixed redshift values and for different band combinations. In each case we quote the mean and the r.m.s. values of $\Delta(log_{10}(sSFR)=log_{10}(sSFR_{output})-log_{10}(sSFR_{input})$.  

\subsubsection{Galaxies simulated with BC03 templates}\label{sec:sSFR_BC3}
For BC03 templates, which are characterized by an exponentially declining star formation histories, the SFR is derived as:
\begin{equation}
SFR_{out}=\frac{e^{\frac{-t_{out}}{\tau_{out}}}}{\tau_{out}}\frac{M_{out}}{M_{0}(t_{out},\tau_{out})}
\label{eq:sSFR_BC03}
\end{equation}
where $t_{out}$, $\tau_{out}$ and M$_{out}$ are the derived values for the age, the characteristic time scale of the declining star formation history and the stellar mass. M$_{0}$ is the mass of the original BC03 template that is used to normalized each template to one solar mass before the SED fitting and it depends on the age and $\tau$ value of each template. The sSFR of BC03 templates depend both on the derived age and star formation history, and not on the derived stellar mass, because $sSFR_{out}=SFR_{out}/M_{out}$. \par
\begin{figure*}[ht!]
\center{
\includegraphics[width=1\linewidth, keepaspectratio]{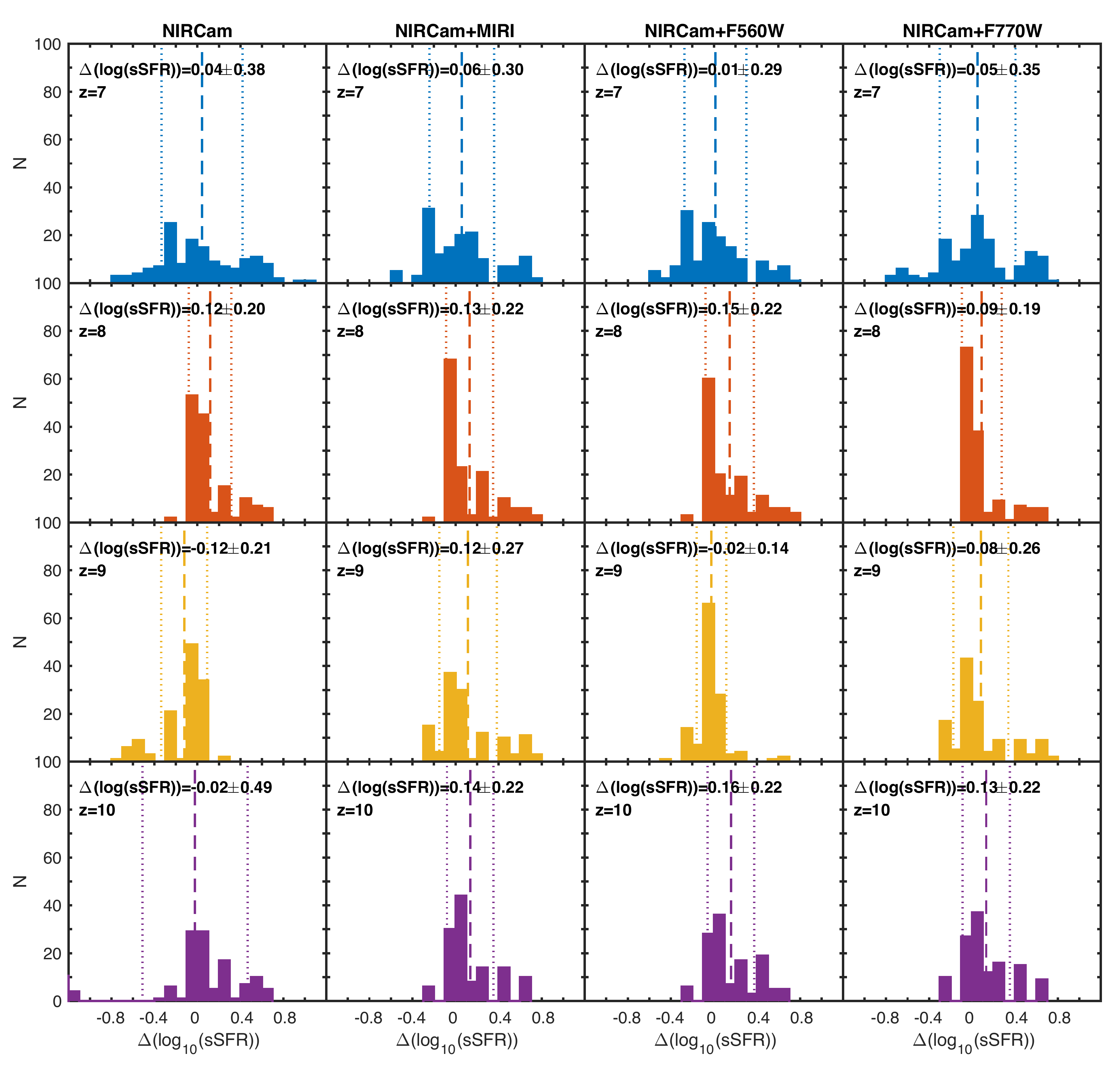}}
\caption{ Differences between the derived and the fiducial sSFR for the BC03 simulated, star-forming galaxies at different fixed redshifts, i.e. $\Delta log_{10}(sSFR)=log_{10}(sSFR_{output})-log_{10}(sSFR_{fiducial})$. \textit{From top to bottom}: redshifts $z=7, 8, 9$ and 10.  sSFRs in each column are obtained with different combinations of \emph{JWST} bands. \textit{From left to right:} 8 NIRCam broad bands; 8 NIRCam broad bands, MIRI F560W and MIRI F770W; 8 NIRCam broad bands and MIRI F560W only; 8 NIRCam broad bands and MIRI F770W only. The vertical lines indicate the mean and the 1 $\sigma_{log(sSFR)}$ values, which are quoted at the top-left of each panel. \label{fig:sSFR_BC03}}
\end{figure*}

The comparison between the original and derived sSFR are shown in figure \ref{fig:sSFR_BC03} for all star-forming galaxies, which correspond to all galaxies with emission lines. Outliers are defined as galaxies with $log_{10}(sSFR_{output})-log_{10}(sSFR_{fiducial})>$3$\sigma_{log(sSFR)}$, with $\sigma_{log(sSFR)}=$0.14 dex, which is the minimum $\sigma_{log(sSFR)}$ obtained with all considered filter combinations. All galaxies simulated with BC03 templates are correctly identified as star-forming at each redshift and for all band combinations. \par
When considering only 8 NIRCam bands, the sSFR difference distribution is broad at z=7 and 10 with $\sigma_{log(sSFR)}=$0.38 dex and 0.49 dex respectively, and it is narrower at the two intermediate redshifts with $\sigma_{log(sSFR)}\sim$0.20 dex. Similarly, the number of outliers is high at z=7 and 10, $\sim34\%$ and 29$\%$ respectively, and it decreases at intermediate redshifts to $\sim15-16\%$. The mode is approximately null at z$\geqslant$8. On the other hand, the mean is around 0 at z$=$7 and 10, but the distribution is more asymmetric at the other two redshifts with $<{log(sSFR)}>=$0.12 dex at z$=$8 and -0.12 dex at z$=$9. The sSFR derivation depends on the age estimation, and this is evident, particularly, at z$=$7 to 9. In particular, at z$=$7 the age difference distribution is broad and therefore the sSFR difference distribution has a similar behavior. 
At z$=$8, the age is always correct or underestimated, so, following equation \ref{eq:sSFR_BC03}, the sSFR difference is only positive or null, while the opposite happens at z$=$9. At z$=$10, the age estimation is better than at z$=$7, because the decreasing age of the Universe limit the age uncertainties, however, no NIRCam band purely covers rest-frame $\lambda>4000\rm\AA$ and the estimation of the star formation history, therefore the characteristic time $\tau$, becomes challenging, as well as the sSFR estimation. \par
When adding the two MIRI bands, the sSFR estimation improves at z$=$7 and, particularly, at z$=$10, where the $\sigma_{log(sSFR)}$ decreases by 0.08 and 0.27 dex, respectively. Similarly, the number of outliers decreases by $\sim40\%$ at these two redshifts. At z$=$7, the MIRI bands improve the age estimation and therefore the sSFR difference distribution is narrower than with only the NIRCam bands. On the other hand, at z$=$10, the MIRI bands improve the $\tau$ estimation covering $\lambda>4000\rm\AA$, that are otherwise purely observed by no NIRCam band. At z$=$8, both the r.m.s and the number of outliers are similar to the values derived with only NIRCam bands, but the number of galaxies with $\Delta log_{10}(sSFR)\sim$0 slightly increases. At z$=$9, both the r.m.s and the number of outliers slightly increase. The age is generally properly estimated at this redshift, but the $\tau$ estimation is very difficult because emission lines are present in all bands at $\lambda>4000\rm\AA$. \par
When considering the two MIRI bands separately, they have similar influence in the sSFR estimation. However, the F560W band reduces the outlier number slightly more at z$=$7 to 9 than the other MIRI band, but none of them decrease the r.m.s. values at all redshifts more than the other. \par
Overall, with only the 8 NIRCam broad-bands, the sSFR estimation is good at z$=$8 and 9. However, it becomes more difficult at z$=$7, because the age estimation is more uncertain, and at z$=$10, because  no NIRCam band purely cover rest-frame $\lambda>4000\rm\AA$ and, therefore, the $\tau$ estimation is challenging. Adding the two MIRI bands improve the sSFR estimation at z$=$7 and 10, but not at intermediate redshifts where it is already good.\par

\subsubsection{Galaxies simulated with \textit{Yggdrasil} templates}\label{sec:SFR_Ygg}
For \textit{Yggdrasil} templates, which are characterized by a step function star formation history, the SFR is derived as:
\begin{equation}
SFR_{out}=\frac{SFR_{constant} M_{out}}{M_{0}(t_{out},SFH)}
\label{eq:sSFR_BC03}
\end{equation}
where SFR$_{constant}$ has different values depending on the durations of the star formation (0.01,0.03 or 0.1 Gyr), M$_{out}$ is the derived stellar mass, while $M_{0}$ is the original mass of the template that it used to normalized each template to 1 $M_{\odot}$ before the SED fitting and it does depend on the SFH and the age. Therefore, for \textit{Yggdrasil} templates, the sSFR, defined as $sSFR_{out}=SFR_{out}/M_{out}$, depend mainly on the SFH and, only maginally, on the age. \par
\begin{figure*}[ht!]
\center{
\includegraphics[width=1\linewidth, keepaspectratio]{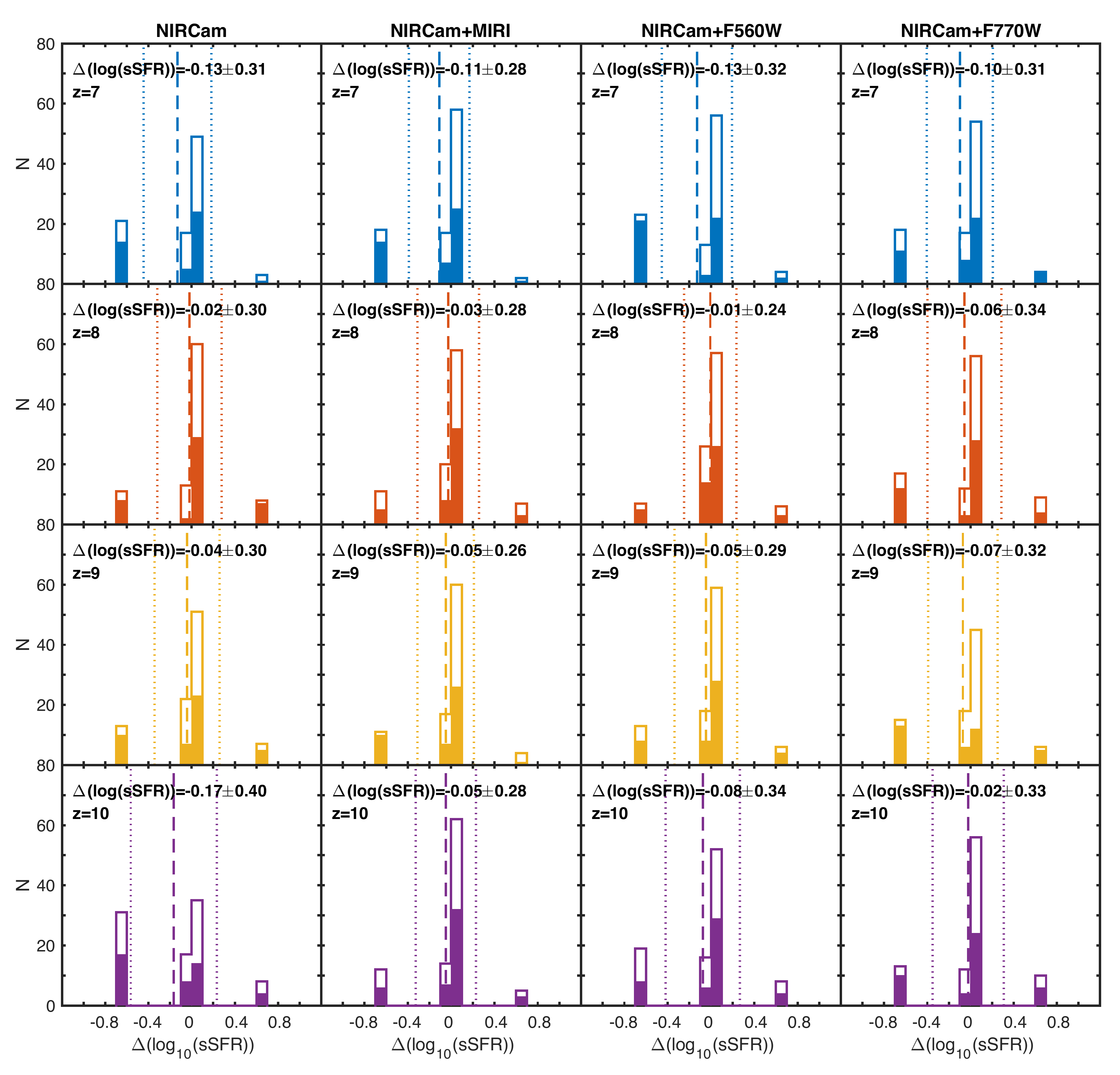}}
\caption{Differences between the derived and the fiducial sSFR for the \emph{Yggdrasil} simulated, star-forming galaxies at different fixed redshifts, i.e. $\Delta log_{10}(sSFR)=log_{10}(sSFR_{output})-log_{10}(sSFR_{fiducial})$.  \textit{From top to bottom}: redshifts $z=7, 8, 9$ and 10.  sSFRs in each column are obtained with different combinations of \emph{JWST} bands. \textit{From left to right:} 8 NIRCam broad bands; 8 NIRCam broad bands, MIRI F560W and MIRI F770W; 8 NIRCam broad bands and MIRI F560W only; 8 NIRCam broad bands and MIRI F770W only. The vertical lines indicate the mean and the 1 $\sigma_{log(sSFR)}$ values, which are quoted at the top-left of each panel. The colored histograms represent galaxies with f$_{cov}=1$, while the white ones represent galaxies with f$_{cov}=0$. \label{fig:sSFR_Ygg}}
\end{figure*}

The comparison between original and derived sSFR are shown in figure \ref{fig:sSFR_Ygg} for all star-forming galaxies. Among \textit{Yggdrasil} templates, there are two types of star-forming galaxies, with and without emission lines, depending on their covering factor. The sSFR depends mainly on the SFH, that has three discrete values, as a consequence, the sSFR difference does not show a smooth distribution, but three separated peaks, a central one and two secondary ones, at all redshifts and with all band combinations. Therefore, we defined outliers as galaxies in the two secondary peaks, i.e. $log_{10}(sSFR_{output})-log_{10}(sSFR_{fiducial})>0.2$ dex. Some star-forming galaxies simulated with \textit{Yggdrasil} templates are wrongly identified as quiescent, i.e. instantaneous SFR$=$0, and their fraction is listed in table \ref{tab:sSFR_Ygg2} for different redshifts and filter combinations. \par
First, we start our analysis considering only the 8 NIRCam bands. Between 3 and 6$\%$ of star-forming galaxies are wrongly identified as quiescent at all redshifts and they are mainly galaxies with f$_{cov}=$1. This is because the presence of numerous emission lines inside all bands can be misidentified as a higher continuum with no emission lines, i.e. a red quiescent galaxy. The r.m.s value of the sSFR difference distribution is around 0.3 dex up to z$=$9 and then it increases at z$=$10 to 0.40 dex. Similarly, the number of outliers is between 20-25$\%$ at z$=$7 to 9 and it increases to 40$\%$ at the highest redshift. Outliers are mainly galaxies with f$_{cov}=$1, indeed the presence of numerous emission lines inside all bands redward the 4000$\rm\AA$ break complicate the identification of the continuum and therefore the estimation of the SFH. Similarly, at z$=$10 no NIRCam band purely covers rest-frame $\lambda>4000\rm\AA$, therefore a proper estimation of the SFH history is difficult. \par
When adding the two MIRI bands, less than 4$\%$ of star-forming galaxies are wrongly identify as quiescent, improving the identification with only  NIRCam bands at all redshifts, but at z$=$9. The r.m.s values are $\sim$0.26-28 at all redshifts, creating a small improvement up to z$=$9 and a more significant one at z$=$10. Similarly, the number of outliers is between 15 and 20$\%$ slightly improving at all redshifts but particularly at z$=$10. As said before, the MIRI bands at z$=$10 can trace $\lambda>4000\rm\AA$ and so it improves the SFH estimation and, as a consequence, the sSFR derivation.\par
When considering each MIRI band separately, with the F560W the fraction of star-forming galaxies wrongly identify as quiescent is smaller than with the F770W. On the other hand, the two bands improve similarly the r.m.s. values and the number of outliers, with no clear preference.\par
Overall, with only observations with the NIRCam bands it is possible to correctly estimate the SFH for $\sim80\%$ of the galaxies up to z$=$9, but the estimation is more difficult at the highest redshift. Adding the two MIRI bands helps to improve the SFH estimation at all redshifts, but particularly at z$=$10.\par 

\begin{deluxetable*}{ccccc}
\tablecaption{Number (percentage) of galaxies among the 384, 96 at each redshift, star-forming galaxies simulated with \textit{Yggdrasil} templates that are wrongly identify as quiescent, for different combination of \emph{JWST} filter combinations and different redshifts. Templates are equally divided between f$_{cov}=$0 and 1 and the number of misidentify galaxies with f$_{cov}=$0 is showed in squared brackets. \label{tab:sSFR_Ygg2}}
\tablecolumns{4}
\tablewidth{0pt}
\tablehead{
\colhead{Bands} &
 \colhead{N$_{outlier,z=7}$} &
 \colhead{N$_{outlier,z=8}$} &
 \colhead{N$_{outlier,z=9}$} &
 \colhead{N$_{outlier,z=10}$} 
}
\startdata
8 NIRCam broad bands &  6 [2] (6.5$\%$) & 4 [2]  (4.2$\%$) & 3 [0]  (3.1$\%$) & 5 [0]  (5.2$\%$)\\
8 NIRCam bands+MIRI F560W, F770W &  1 [0]  (1.0$\%$) & 0 [0] (0.0$\%$) & 4 [0] (4.2$\%$) & 3 [3] (3.1$\%$)\\
8 NIRCam bands+MIRI F560W & 0 [0] (0.0$\%$) & 0 [0] (0.0$\%$) & 4 [0] (4.2$\%$) & 3 [0] (3.1$\%$)\\
8 NIRCam bands+MIRI F770W  &  3 [0] (3.1$\%$) & 2 [1] (2.1$\%$) & 12 [0] (12.5$\%$) & 5 [1] (5.2$\%$)\\
\enddata
\end{deluxetable*}

\section{Summary and conclusions}\label{sec:conclusions}
In this work we have tested the impact of having data in different \emph{JWST} filter combinations on deriving stellar mass, age, color excess and sSFR for a sample of simulated galaxies at $z=7-10$. In particular, we considered the 8 NIRCam broad-band filters and the two MIRI filters at the shortest wavelengths (F560W and F770W), which are the most sensitive ones among all MIRI filters and those which will be preferred for high-z galaxy surveys.  \par
Our sample consists of 1542 simulated galaxies from B16, derived from BC03 templates with manual addition of emission lines for star-forming galaxies (which are the ones with age lower than the characteristic time $\tau$ of the star formation history) and \emph{Yggdrasil} models. All galaxies in our sample have good photometric redshift estimations, as derived in B16, therefore uncertainties on the photometric redshifts are not a significant source of error on the galaxy properties estimation performed here. The sample contained the typical SED types of the galaxies that will be observed in \emph{JWST} NIRCam and MIRI high-z blank survey, in order to test potential problems on deriving galaxy properties, on an equal basis. Therefore, the used sample does not try to emulate the real distribution of galaxies at high redshift.\par
Our main results are:\par
$\bullet$ For galaxies with conventional SED types (i.e. BC03 templates), with the presence of at most some prominent emission lines, stellar masses can be well recovered with NIRCam broad-band data alone up to $z=9$, provided that these data have sufficiently high S/N values ($>10$). At $z=10$,  no NIRCam filter purely cover rest-frame $\lambda>$4000$\rm\AA$, therefore the stellar mass estimation is more difficult, resulting on 31.1$\%$ of 3$\sigma_{logM^{*}}$ outliers. For the same reason, the MIRI bands have an important role in improving the mass estimation at this redshift, reducing the number of outliers to $\sim13\%$. \par
$\bullet$ When templates with rest-frame equivalent widths that evolve with redshift are considered, it is preferable to derive stellar masses in two steps, in order to avoid the code to use a template with rest-frame equivalent widths of different redshifts. First, one should derive the photometric redshift, that is virtually not affected by this problem, and after that, derive the stellar mass at the fixed obtained redshift. \par
$\bullet$ For galaxies with nebular continuum and emission, the stellar mass is generally overestimated when considering only NIRCam broad-band observations. The incorporation of MIRI bands is fundamental for a proper stellar mass recovery, except at $z=9$, where the stellar mass recovery remains anyway challenging. The reasons are that, first, the number of NIRCam bands at $\lambda>4000\rm\AA$ decreases with redshift and, second, emission lines contaminate these bands. In particular, at $z=9$, emission lines are inside all NIRCam and MIRI bands at $\lambda>4000\rm\AA$ and it is not possible to set the level of the continuum, making the mass estimation challenging. \par
$\bullet$ The NIRCam broad-band filters are sufficient to recover galaxy age for $>64\%$ of the sample within 0.03 Gyr, and to recover the color excess for $>70\%$ of the sample within 0.06 mag. However, extreme outliers are present at all redshifts, particularly at $z<9$. Indeed, at higher redshifts, thanks to the decreasing age of the Universe, the age-dust degeneracy is reduced.\par
$\bullet$ The two MIRI bands improve both the age estimation and the color excess estimation, slightly decreasing the r.m.s. values, but mainly reducing the number of outliers. \par
$\bullet$ With observations in the 8 NIRCam broad-band filters, the sSFR estimation is challenging at z$=$10, where  no NIRCam band purely cover $\lambda>4000\rm\AA$, so adding MIRI observations improves the sSFR derivation. At lower redshifts, NIRCam alone is generally sufficient to recover the sSFR for $\sim70-80\%$ of star-forming galaxies within 0.4 dex.\par 
The results presented in this paper are a useful reference for the design of deep imaging surveys with \emph{JWST} and to understand in which situation the time investment on MIRI observations is necessary to recover galaxy properties, i.e. stellar masses at the highest redshift. Indeed, for the majority of $7\leqslant z\leqslant 9$ galaxies, it is possible to estimate the stellar mass, the sSFR, the age and the color excess at all redshifts just using 8 NIRCam bands. However, adding the MIRI bands, both F560W and F770W, helps improve the estimation of these quantities, and it is essential to recover the stellar mass and the sSFR of galaxies at z$=10$. On the other hand, the stellar mass of young star-forming galaxies with emission lines ($f_{cov}>0$) remain challenging at any redshift. Overall, it will be necessary to observe with both cameras in order to perform an accurate study of galaxy evolution and mass assembly since early cosmic times.

\acknowledgments
LB and KIC acknowledge the support of the Nederlandse Onderzoekschool voor de Astronomie (NOVA).  KIC also acknowledges funding from the European Research Council through the award of the Consolidator Grant ID 681627-BUILDUP. OLF acknowledges funding from the European Research Council Advanced Grant ID 268107-EARLY.  PGP-G acknowledges support from the Spanish Government MINECO Grants AYA2012-31277 and AYA2015-70815-ERC. LC acknowledges support from the Spanish Government MINECO Grants AAYA2012-32295. 

\vspace{5mm}




\appendix

\section{Increasing star formation histories}\label{sec:AppA}

Recent simulations of the high redshift Universe predict that galaxies at z$>$6 have star formation rates that, on average, increase with time \citep{Finlator2011,Jaacks2012,Zackrisson2017}. Following these predictions, we tested the stellar mass derivation for a set of templates with a single increasing SFH. In particular, we considered a BC03 template with a delayed star formation history with a $\tau$ value of 1 Gyr, for which the SFR increases up to 1 Gyr and then exponential decline. We considered solar metallicity and the same extinction and age values used for the other BC03 templates. In particular, all age used are below 1 Gyr so all templates analysed in this appendix have rising SFH.  We manually added emission lines as for the other BC03 templates. We derived mock photometry for the JWST bands and the output parameters, fixing the redshift, as it is described in this paper for the other BC03 SED templates. \par
\begin{figure}[ht!]
\center{
\includegraphics[width=1\linewidth, keepaspectratio]{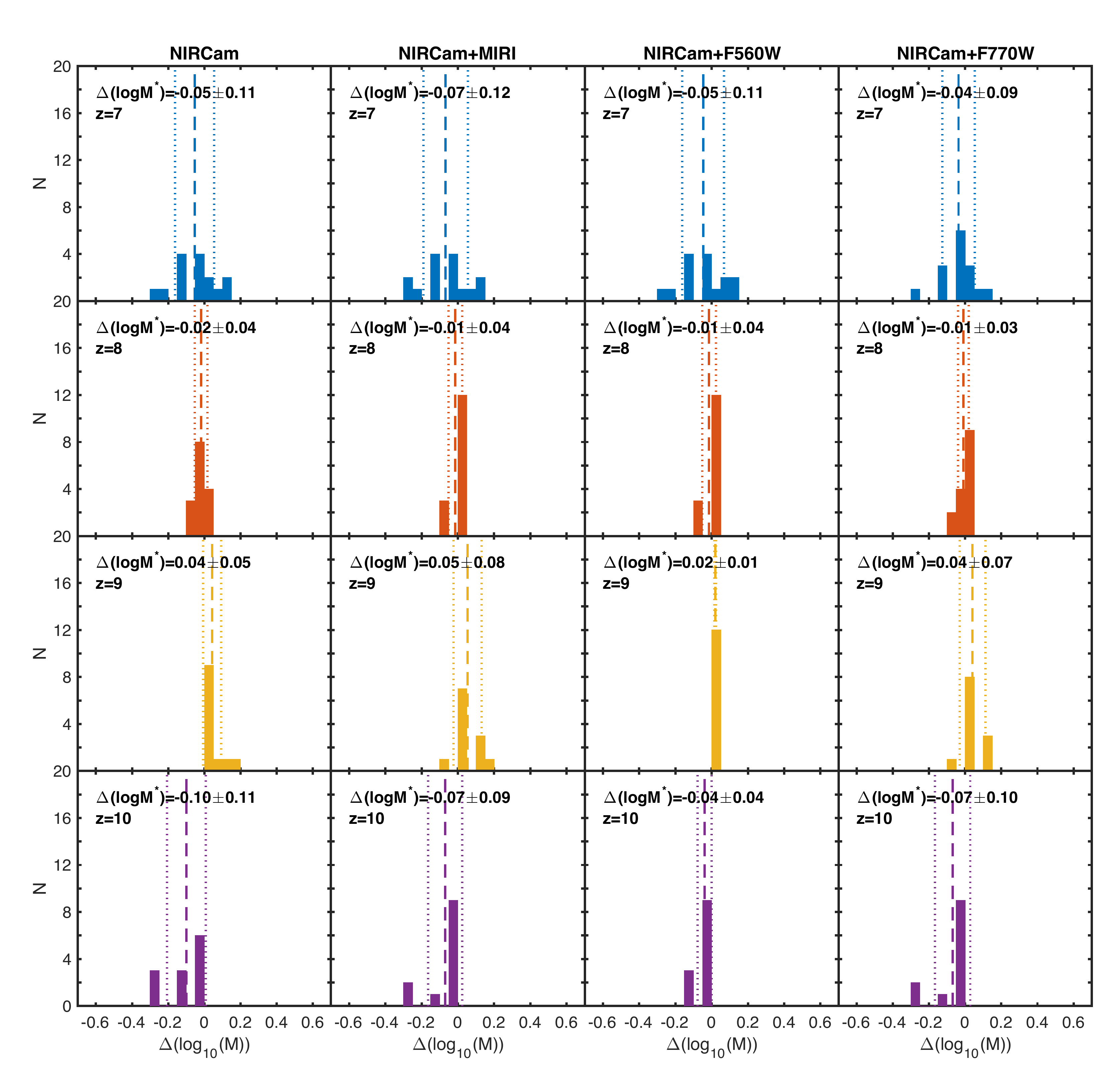}}
\caption{ Differences between the derived and the fiducial stellar mass for the BC03 simulated galaxies with delayed SFH at different fixed redshifts, i.e. $\Delta log_{10}(M^{*})=log_{10}(M^{*}_{output})-log_{10}(M^{*}_{fiducial})$.  \textit{From top to bottom}: redshifts $z=7, 8, 9$ and 10.  Stellar masses in each column are obtained with different combinations of \emph{JWST} bands. \textit{From left to right:} 8 NIRCam broad bands; 8 NIRCam broad bands, MIRI F560W and MIRI F770W; 8 NIRCam broad bands and MIRI F560W only; 8 NIRCam broad bands and MIRI F770W only. The vertical lines indicate the mean and the 1 $\sigma_{logM^{*}}$ values, which are quoted at the top-left of each panel. \label{fig:mass_del}}
\end{figure}

The distribution of the difference between the derived stellar mass and the fiducial ones is shown in Figure \ref{fig:mass_del}. With only 8 NIRCam bands the stellar mass estimation is good at z$=$8-9, while the r.m.s. is 0.11 at the other redshifts. Adding the two MIRI bands mainly improves the stellar mass estimation at z$=$10, which is otherwise generally underestimated with only 8 NIRCam bands. The F560W generally improves more the stellar mass recovery respect to the F770W. \par
This simple test does not take into account possible degeneracies among delayed star formation histories with different $\tau$ values or different metallicities. However, as for the other BC03 templates with emission lines, the stellar masses are generally correctly derived with 8 NIRCam bands at z$=$7-9 and the inclusion of MIRI bands is necessary to improve the recovery of the stellar mass at z$=$10.

\end{document}